\documentclass{article}

\usepackage{graphicx} 
\usepackage[hidelinks]{hyperref}
\usepackage{cleveref}
\usepackage{multicol}
\usepackage{tikz}
\usepackage[font=small]{caption}
\usetikzlibrary{positioning, arrows.meta, calc, shapes.geometric}

\newcommand{\full}{\tikz[baseline=-0.6ex]\fill[black!75] (0,0) circle (0.13);}
\newcommand{\half}{\tikz[baseline=-0.6ex]{\draw[black!75] (0,0) circle (0.13);
  \fill[black!75] (0,0) -- (90:0.13) arc (90:270:0.13) -- cycle;}}
\newcommand{\none}{\tikz[baseline=-0.6ex]\draw[black!55] (0,0) circle (0.13);}

\usepackage{authblk}
\usepackage[
backend=biber,
style=apa,
sorting=nyt
]{biblatex}

\title{Interactive Communication - cross-disciplinary perspectives from psychology, acoustics, and technology}

\author[1]{Mareike Daeglau*}
\author[2]{Stephan Getzmann*}
\author[3]{Moritz Bender}
\author[4]{Janina Fels}
\author[5]{Rainer Martin}
\author[6]{Alexander Raake}
\author[7]{Isabel S. Schiller}
\author[7]{Sabine J. Schlittmeier}
\author[8]{Katrin Schoenenberg}
\author[9]{Felix Stärz}
\author[10]{Leon O. H. Kroczek}

\affil[1]{Neuropsychology Lab, Department of Psychology, Carl von Ossietzky University of Oldenburg,
Oldenburg, Germany}
\affil[2]{Department of Ergonomics, IfADo - Leibniz Research Centre for Working Environment and Human Factors at TU Dortmund, Dortmund, Germany}
\affil[3]{Department of Medical Physics and Acoustics, Carl von Ossietzky University of Oldenburg, Oldenburg, Germany}
\affil[4]{Institute for Hearing Technology and Acoustics, RWTH Aachen University, Aachen, Germany}
\affil[5]{Institute of Communication Acoustics, Ruhr-Universit{\"a}t Bochum, Bochum, Germany}
\affil[6]{Chair and Institute of Communications Engineering, RWTH Aachen University, Aachen, Germany}
\affil[7]{Work and Engineering Psychology, RWTH Aachen University, Germany, Aachen, Germany}
\affil[8]{Department of Psychology, Clinical Psychology and Psychotherapy, Kiel University, Kiel, Germany}
\affil[9]{Institute of Hearing Technology and Audiology, Jade University of Applied Sciences, Oldenburg, Germany}
\affil[10]{Department of Psychology, Clinical Psychology and Psychotherapy, University of Regensburg, Regensburg, Germany}

\date{}

\begin{document}

\maketitle
* Equal contribution\\

Corresponding author:
Leon O. H. Kroczek, Mail: leon.kroczek@ur.de

\newpage



\begin{abstract}

    Interactive communication (IC), i.e., the reciprocal exchange of information between two or more interactive partners, is a fundamental part of human nature. As such, it has been studied across multiple scientific disciplines with different goals and methods. This article provides a cross-disciplinary and selective primer on contemporary IC integrating psychological mechanisms with speech signal, acoustic and media-technological constraints in theory, measurement, and applications. First, we outline theoretical frameworks that account for verbal, nonverbal, and multimodal aspects of IC, including distinctions between face-to-face and computer-mediated communication. Second, we summarize key methodological approaches, including behavioral, cognitive, and experiential measures of communicative synchrony and acoustic signal quality. Third, we discuss selected applications, applications in which speech transmission, signal enhancement, mediated dialogue, and real-time coordination are central, namely assistive listening technologies, conversational agents, and social VR, alongside ethical considerations. Taken together, this primer highlights how human capacities and technical systems jointly shape IC, consolidating concepts, findings, and challenges that have often been discussed in separate lines of research. 
    
\end{abstract}

\newpage


\section{Motivation and Goal}
In recent decades, interactive communication (IC) has shifted from predominantly face-to-face encounters to a spectrum of technology-mediated formats, from telephone and videoconferencing to virtual and extended reality (VR, XR). This shift changes the cues, timing, and coordination an exchange can rely on, and it does so first of all at the level of the speech signal, which remains the main channel through which people interact in real time \parencite{walther1996computer}. Communication science has theorized this move with frameworks such as Media Richness Theory \parencite{DaftLengel} and Social Presence Theory \parencite{PsychTelecomm}, which describe how media differ in cue availability and immediacy. These frameworks, however, were developed before the acoustic and signal-processing properties of mediated speech became a central design variable, and they say little about how the speech signal itself shapes interaction.

Some of these formats have been used for a long time, and their effects are well established, for instance in telephony and video conferencing. Even there, the main differences from real-life face-to-face communication come from an impoverished presentation of interactive cues, especially the acoustic ones (cf. \cite{skowronek2022quality, Raake2022, eddy2019technology}). Such reductions in visual and auditory feedback alter perceived immediacy, empathy, and coordination efficiency in online meetings \parencite{FAUVILLE2021100119,Bailenson2021Nonverbal}, and adverse acoustic conditions like noise, reverberation, or latency make speech harder to follow just when timing matters most \parencite{pichora2015hearing}. In contrast, VR can restore a richer set of cues such as body posture, gestures, and gaze, and, when latency, tracking fidelity, spatial audio, and avatar expressivity are sufficient, it may approach face-to-face conditions.

 While IC depends both on what people can perceive and process and on what the technology allows the speech signal to carry, the two sides have mostly been studied separately, in psychology on one side and in acoustics and communication technology on the other. Therefore, we argue in this primer that they are best analyzed together, with the speech signal as the common ground: it is what psychological coordination works on, and what a technical system either preserves or degrades. This is most relevant for VR-mediated IC, which is still emerging and where the interplay of verbal and nonverbal information is both easier to control and more easily disturbed by the technology than in established media.

This is a selective, cross-disciplinary primer rather than a systematic review, and we make no claim to cover the field exhaustively; instead we draw on work from psychology, acoustics, and communication technology that shows how these two sides interact in spoken, multimodal, and technology-mediated communication. We prioritize work that examines how cognitive, social, and perceptual processes unfold through speech, acoustic, and multimodal signals, and how these signals are shaped by technological mediation in communication environments. On this basis we contribute by (1) defining IC via bidirectionality, contingency, mutual awareness, and temporal coupling, criteria that align with interactionist definitions of dialogue \parencite{Clark1996, Pickering_Garrod_2004}; (2) integrating psychological and media-technological theory into a common framework centered on the speech and acoustic signal and its technological mediation (Figure~\ref{fig:framework}); (3) organizing measurement approaches by signal quality, synchrony, and experience; and (4) mapping applications such as assistive listening, conversational agents, and social VR, together with the ethical questions they raise.

\section{Definition}\label{Definition}

The term IC is frequently used in psychology, communication science, human-computer interaction, and other fields, but the different disciplines may vary in their understanding of what IC actually means. Psychological accounts, for instance, foreground the cognitive and social mechanisms that let partners coordinate, such as turn-taking, mutual prediction, and the building of common ground, whereas acoustic and communication-technological accounts foreground the channel itself and how faithfully it carries the signals those mechanisms depend on. In this cross-disciplinary primer we therefore synthesize these differing understandings into one inclusive definition of IC that can be applied across disciplines. Here, IC describes the bidirectional and dynamic process of transferring information between two or more interactive partners. A key feature of IC is a contingent feedback loop: received information is used to generate a response that is time- and content-contingent and dependent on the partner. Operationally, IC requires (i) bidirectionality, (ii) response contingency, (iii) mutual awareness, and (iv) temporal coupling; these criteria distinguish IC from broadcast or unlinked, asynchronous exchanges, and they are meant as a common denominator across the disciplinary perspectives in Section~\ref{Theory} rather than as a definition tied to any single field. Importantly, the transfer of information can comprise verbal and nonverbal channels in different sensory modalities \parencite{sebeok2001signs}. Verbal information can be presented in the auditory (vocal and verbal: spoken language) or visual domain (non-vocal and verbal: sign language). Similarly, nonverbal information can also include the auditory (vocal and nonverbal: e.g., prosody) and visual domain (non-vocal and nonverbal: e.g., gaze, facial expressions, gestures).

The following real-life situations illustrate this definition. First, imagine a lecturer in front of an audience. The lecturer is the only person speaking, there are no questions/comments from the audience. Does this still qualify as IC? We argue that the answer is yes, even though only one person is speaking. Yet, the (nonverbal) behavior of the audience will communicate something to the lecturer. Audience members may establish eye contact or nod, signaling that they can follow the explanations, or they may look away or display a puzzled expression, indicating that the lecturer should adjust the pace or provide clarification. Importantly, this applies not only to face-to-face situations but also to virtual meetings via video calls, where fewer communicative channels are available (e.g., delayed responses, restricted gaze cues). In terms of our criteria, the lecture satisfies mutual awareness and contingency via gaze and nods, and temporal coupling within the shared setting. Research on nonverbal feedback in teaching contexts confirms that cues such as gaze, nodding, and posture strongly influence perceived engagement and comprehension \parencite{Kleinke1986}.

As a second scenario, consider a situation at a train station. You are waiting for your train to arrive and a loudspeaker informs you that the train will be delayed by 30 minutes. You respond with an angry expression and exclaim, "How can it be so difficult to simply be on time?". Does this constitute IC? We argue it does not, even though both parties express something. Here, mutual awareness and contingency are absent; thus, the exchange remains unidirectional.

However, some situations make it difficult to draw a clear line between unidirectional and interactive communication. For example, consider interacting with a chatbot or virtual agent in a VR application (in this work the term agent is used to distinguish a computer-controlled virtual character from human-controlled character, i.e. an avatar). Although such entities are not real humans, they can be programmed to produce responses that closely mimic, or even become indistinguishable from those of a human interlocutor. Advanced Artificial Intelligence (AI)-based designs enable a dynamic exchange that transcends one-way communication and engages the user in a responsive and interactive manner. Consequently, we argue that such interactions can be considered IC, even though they involve an artificial agent. This interpretation is consistent with the ‘Computers as Social Actors’ paradigm, which demonstrates that humans tend to apply social norms to responsive technologies \parencite{nassMachinesMindlessnessSocial2000,reeves1996media}.

\section{Theory}\label{Theory}
Understanding IC requires theoretical perspectives that account for both the psychological mechanisms underlying communication and the technological conditions shaping it. In this section, we outline two complementary viewpoints: (1) the psychological perspective, focusing on the modalities and processes of human communication, and (2) the technological and acoustic perspective, examining how modalities and technical systems shape the affordances and constraints of communication. By integrating these perspectives, we argue that IC should be analyzed not only in terms of signal exchange, but also as a functionally embedded, socially co-constructed, and technologically mediated activity. Accordingly, theoretical integration must address how cognitive mechanisms interact with the affordances and constraints of communication media systems to sustain coordination and shared understanding. For example, gaze-based turn-yielding relies on psychological prediction and on technical conditions (e.g., video frame rate and AV-sync) in mediated settings. Figure~\ref{fig:framework} summarizes this integrative view by placing the shared speech and acoustic signal at the center of the interaction and showing how technical mediation can preserve, transform, or degrade this shared substrate.

\begin{figure}[t]
\centering
\resizebox{\textwidth}{!}{%
\begin{tikzpicture}[
  font=\footnotesize,
  >={Latex[length=2.4mm]},
  head/.style={draw, circle, minimum size=1.0cm, fill=black!6, line width=0.5pt},
  body/.style={draw, trapezium, trapezium left angle=68, trapezium right angle=68,
               minimum width=1.7cm, minimum height=0.85cm, fill=black!6, line width=0.5pt},
  chan/.style={draw, rounded corners, align=center, fill=black!10, text width=5.4cm,
               minimum height=1.7cm, thick},
  band/.style={draw, rounded corners, align=center, fill=black!3, text width=12.8cm, minimum height=0.8cm},
  lab/.style={align=center},
]
\node[head] (hA) at (0,0) {};
\node[body, below=0.04cm of hA] (bA) {};
\node[lab, font=\footnotesize] at (0,-1.95) {\textbf{Person A}};
\node[head] (hB) at (10,0) {};
\node[body, below=0.04cm of hB] (bB) {};
\node[lab, font=\footnotesize] at (10,-1.95) {\textbf{Person B}};
\node[chan] (ch) at (5,0) {\textbf{Speech \& acoustic signal}\\[1pt]{\scriptsize directly (face-to-face) or via a\\ technical channel: latency, noise,\\ bandwidth, spatial audio}};
\draw[<->, thick] (hA.east) -- node[above, align=center, font=\scriptsize]{verbal $+$\\nonverbal} (ch.west);
\draw[<->, thick] (ch.east) -- node[above, align=center, font=\scriptsize]{verbal $+$\\nonverbal} (hB.west);
\draw[<->] (hA.north) to[bend left=24]
  node[above, align=center, yshift=1pt]
  {\textbf{interactive communication}\\bidirectional $\cdot$ contingent $\cdot$ mutually aware $\cdot$ temporally coupled}
  (hB.north);
\node[band] (m) at (5,-2.85) {\textbf{Assessed via:} signal quality $\cdot$ synchrony $\cdot$ experience};
\node[lab, font=\footnotesize] at (5,-3.75) {Contexts (e.g.): face-to-face $\cdot$ video $\cdot$ VR $\cdot$ assistive listening $\cdot$ conversational agents};
\end{tikzpicture}%
}
\caption{An interaction-centred framework for interactive communication (IC). Two partners exchange verbal cues (speech) and nonverbal cues (gaze, facial expression, gesture, prosody) through a shared speech and acoustic signal, either directly in a face-to-face setting or by means of a technical channel, which can preserve or degrade the signal, for instance through latency, noise, limited bandwidth, or the rendering of spatial audio. The exchange is bidirectional, contingent, mutually aware, and temporally coupled, the four criteria by which we define IC in Section~\ref{Definition}. A technical channel can degrade the signal and, with it, these criteria, for instance when latency reduces temporal coupling. The resulting interaction can be assessed along families of measures such as signal quality, synchrony, and experience (Section~\ref{Measures}), and the same structure recurs across contexts such as face-to-face conversation, video conferencing, VR, assistive listening, and conversational agents (Section~\ref{Applications}). The figure is a schematic illustration of the concepts discussed in this primer, not a comprehensive overview of the field of interactive communication; further measures and contexts are possible.}
\label{fig:framework}
\end{figure}

\subsection{A Psychological Perspective on IC}\label{Psychology}
From a psychological perspective, IC can be analyzed with respect to the perceptual and expressive modalities through which information is exchanged. In this section, we organize psychological perspectives by modality, verbal, nonverbal, and multimodal communication, highlighting the underlying cognitive, social, and emotional processes, as well as contextual and technological influences. Across these aspects, IC is fundamentally adaptive: interlocutors continuously interpret and adjust to each other’s verbal and nonverbal signals to establish, maintain, and repair coordination. 

IC is not merely the exchange of information; it is a goal-directed, dynamically coordinated activity engaging cognitive, social, and affective processes. Traditionally, IC has been described as a function of social interaction, namely the transfer of information from one person to another (\cite{hadleyReviewTheoriesMethods2022}). As such, it has been differentiated from other functions, such as affiliation and social cognition (\cite{Frith2012}). Importantly, however, affiliation and social cognition also rely on decoding communicative signals (e.g., a smile, direct gaze) to establish rapport, infer intentions, and generate social evaluations (\cite{argyleBodilyCommunication2013}). IC can therefore be regarded as an important source of information for social processes, such as fostering affiliation and establishing social hierarchies.

The following sections examine how key modalities contribute to IC, starting with verbal (vocal) information exchange under adverse acoustic conditions, moving to the role of nonverbal (vocal and nonvocal) signals in social inference and coordination, and concluding with their integration in multimodal settings. We then consider how communication changes across media environments and over the lifespan. Together, these sections illustrate how IC is shaped by human capacities, social functions, and contextual constraints.
 
\subsubsection{The Verbal Modality and Effects on Acoustic Interference}
To begin, we discuss the verbal modality, as it is oftentimes considered the primary channel for information exchange. A core challenge for verbal communication arises when acoustic signals are degraded, whether due to environmental noise, room effects such as reverberation, or the speaker’s voice characteristics. Such degradation can impede effective communication and increase cognitive load, especially in interactive settings. The most widely studied acoustic challenge is noise, which can originate from environmental sources (e.g., construction, classroom, or ventilation noise) or from competing speakers (e.g., irrelevant background speech). Both types of noise can significantly impair cognitive performance (\cite{Marsh2023}), as they reduce the resources available for core demands of IC, such as speech perception, speech production, integration into long-term memory, and interpretation of turn-taking cues. Studies demonstrate that adverse signal-to-noise ratios (SNRs) significantly impair dialogue comprehension, especially when interactional timing is critical \parencite{pichora2015hearing}. In mediated communication, additional noise may be introduced by the system; conversely, algorithms may process or cancel environmental noise at sender or receiver. Speakers adapt via the Lombard effect (i.e., increased level and spectral shifts, \cite{lane1971,lombard1911}), which interacts with intelligibility and listening effort in IC.

Room effects such as reverberation can also shape communicative success, with and without background noise. Poor room acoustics can impair speech intelligibility and overall communication quality, necessitating careful consideration of these factors in interactive settings (\cite{Ermert2025b, Seitz2024, Yadav2023}).

Beyond noise, speech may also be degraded when the communication partner has a voice disorder, most commonly characterized by hoarseness or dysarthritic speech. Besides people affected by neurodegenerative diseases, voice disorders are particularly prevalent among professionals who rely heavily on their voice, affecting, for example, 41\% of university professors (\cite{Azari2024}). Research indicates that listening to a hoarse speaker requires greater listening effort and is perceived as more annoying, while also impairing cognitive performance (\cite{Schiller2023, Schiller2024}). In IC, where understanding and responding are tightly interlinked, noise or voice impairment can therefore increase cognitive effort, reduce speech intelligibility and comprehension, and hinder the coordination of the interaction. Another key factor is the SNR and absolute level, which interact with voice quality and background noise to shape intelligibility. In mediated communication, voice degradations may also result from speech coding or transmission errors. Technology-induced effects are discussed in Section \ref{TechnologyPerspective}.

\subsubsection{The Nonverbal Modality and Social Cognition}\label{NonverbalModalitySocialCognition}
Although many psychological mechanisms of communication operate across modalities, they are particularly evident and empirically tractable in nonverbal channels. Social cues such as gaze, facial expressions, posture, vocal prosody and other nonverbal vocalizations are critical for regulating interaction and conveying social intent. They provide meta-communicative information (e.g., about turn-taking, attention, or affective stance) and help recipients interpret and disambiguate verbal content. The interpretation of such cues depends on shared conventions and situational context, and mismatches in their production or perception can disrupt interaction and hinder mutual understanding. In line with Conversation Analysis (\cite{sacks}), nonverbal timing cues are essential for regulating speaker changes and avoiding overlap, thereby maintaining conversational flow.

Nonverbal communication engages cognitive processes involved in interpreting communicative signals. Nonverbal behaviors also serve a predictive function: interlocutors continuously anticipate others’ reactions based on micro-movements and gaze, facilitating rapid adaptation. The Emotion-as-Social Information (EASI) model (\cite{langeReadingEmotionsReading2022,vankleefHowEmotionsRegulate2009}) suggests that individuals use emotional expressions (e.g., facial emotional expressions, but also other signals like body posture or voice) of an interaction partner to infer their state and intentions. Ultimately, any perceivable behavior, even a lack of response, may be interpreted as meaningful, in line with Paul Watzlawick's first axiom that "one cannot not communicate" (\cite{watzlawick2017tentative}). More broadly, these inferential processes extend across modalities but become especially salient in nonverbal channels, and are captured under the concept of "Theory of Mind" (ToM; \cite{frithTheoryMind2005}). Here, inferential processes enable individuals to assess whether an interaction partner shares relevant knowledge or whether additional information must be conveyed, a capacity central to establishing common ground (\cite{danielc.richardsonArtConversationCoordination2007}). Taken together, these mechanisms underscore the interpretive nature of communication and the foundational role of nonverbal cues in achieving shared understanding.

\subsubsection{Integrating Verbal and Nonverbal Information}\label{IntegratingVerbalNonverbalInformation}
In face-to-face interaction, verbal and nonverbal behaviors are closely intertwined. Each channel (e.g., acoustic, visual, tactile) can convey information independently, for example, giving verbal instructions while maintaining eye contact to check attentiveness. However, in many situations verbal and nonverbal signals must be integrated to decode the intended message, such as when speech is ambiguous or when irony is used (\cite{Holle2007}). In addition, nonverbal channels facilitate coordination between interactive partners. For instance, eye gaze is used to signal turn endings in face-to-face conversations, enabling smooth and rapid interactions (\cite{wohltjenEyeContactMarks2021}). Similar mechanisms have been described for body posture (\cite{matsumotoBodyPosturesGait2016,romero2021visual}) and manual gestures (\cite{kendrickTurntakingHumanFacetoface2023}), confirming the multimodal nature of turn-taking in human face-to-face conversations. These integrative mechanisms are especially critical when information is impaired, incomplete, or not redundantly conveyed—as often occurs in video-mediated communication  (\cite{bohannonEyeContactVideomediated2013}). Classic demonstrations such as the McGurk effect (\cite{mcgurkHearingLipsSeeing1976}) underscore obligatory audio-visual integration. While recent evidence suggests that audio-visual incongruence can be ignored if it occurs in a domain irrelevant to the experimental task (\cite{Ermert2025a}), further enhancing the understanding of the interplay of verbal and nonverbal cues is essential for explaining how communicative coordination is maintained, or breaks down, especially when IC is technologically mediated or constrained.

\subsubsection{Face-to-Face versus Computer-Mediated Communication} \label{F2FvCMC}
Face-to-face and computer-mediated communication (CMC) differ in several psychological aspects (\cite{tsigeman2024psychological}). Face-to-face communication benefits from nonverbal cues such as body language and tone of voice, which enhance emotional expression, empathy, and immediacy of feedback, thereby fostering deeper connections. CMC often lacks these cues, which can increase misunderstandings and cognitive load as interlocutors must interpret messages without immediate clarification. Face-to-face interactions are typically more persuasive and effective in changing attitudes due to the richness of real-time engagement. CMC, by contrast, allows greater control over self-presentation. While this can be advantageous, it may also encourage inauthentic communication. Overall, face-to-face communication is richer and more immediate, whereas CMC offers flexibility but often at the expense of connection and emotional depth. These contrasts align with Media Synchronicity Theory (\cite{dennis}), which posits that differences in transmission speed and feedback capability explain much of the reduced sense of co-presence in CMC. However, CMC is highly diverse (\cite{yao2020computer}), and novel VR-based forms are continuously being developed. Thus, some existing disadvantages may diminish further in the future. This comparative perspective highlights how communicative modality interacts with psychological mechanisms and underscores the importance of analyzing medium-specific demands, constraints, and opportunities.

\begin{figure}[ht]
\centering
\resizebox{\textwidth}{!}{%
\begin{tikzpicture}[font=\footnotesize]
\def\colsep{2.5}
\def\rowsep{0.72}
\foreach \i/\name in {0/{Face-to-face}, 1/{Telephone}, 2/{Video conf.}, 3/{Social VR}, 4/{Conversational\\ agent}}
  \node[align=center, font=\scriptsize\bfseries] at (\i*\colsep, 0.95) {\name};
\foreach \j/\rowname/\a/\b/\c/\d/\e in {%
  0/{Verbal content (speech)}/\full/\full/\full/\full/\full,
  1/{Prosody, vocal timing}/\full/\half/\half/\half/\half,
  2/{Spatial audio cues}/\full/\none/\none/\half/\none,
  3/{Gaze direction}/\full/\none/\half/\half/\half,
  4/{Facial expression}/\full/\none/\full/\half/\half,
  5/{Gesture, posture}/\full/\none/\half/\half/\half,
  6/{Temporal contingency}/\full/\half/\half/\half/\half%
}{
  \node[anchor=east, font=\scriptsize] at (-1.15, -\j*\rowsep) {\rowname};
  \node at (0*\colsep, -\j*\rowsep) {\a};
  \node at (1*\colsep, -\j*\rowsep) {\b};
  \node at (2*\colsep, -\j*\rowsep) {\c};
  \node at (3*\colsep, -\j*\rowsep) {\d};
  \node at (4*\colsep, -\j*\rowsep) {\e};
}
\draw[black!45] (-1.05, 0.5) -- (4*\colsep+0.55, 0.5);
\draw[black!45] (-1.05, -6*\rowsep-0.42) -- (4*\colsep+0.55, -6*\rowsep-0.42);
\node[anchor=west, font=\scriptsize] at (-1.05, -6*\rowsep-0.95)
  {\full\, preserved \quad \half\, degraded or partial \quad \none\, absent};
\end{tikzpicture}%
}
\caption{Schematic comparison of how selected communication settings preserve, degrade, or remove the cues on which interactive communication depends. Face-to-face interaction serves as the reference case in which all cues are available. Verbal content survives in every setting, whereas nonverbal and spatial cues are lost or attenuated to different degrees, and social VR partially restores cues that telephony and video conferencing reduce. For conversational agents, the availability of nonverbal cues such as gaze, facial expression, and gesture depends on whether the agent is embodied. Temporal contingency is shown as the enabling condition under which the remaining cues can be exploited for coordination, rather than as a cue in itself. The assignments are schematic and typical rather than absolute, since they depend on the specific implementation, for instance on whether a video system transmits spatialised audio or whether an agent supports barge-in. They summarise the evidence discussed in Sections~\ref{Theory} and~\ref{Applications}.}
\label{fig:cuematrix}
\end{figure}

\subsubsection{Developmental Variation and Age Aspects} \label{Age}
Age is an important factor shaping how individuals engage in IC across contexts and modalities. Older adults often draw on their life experience, which can enhance communication skills, particularly in face-to-face settings, where they may be more adept at reading nonverbal cues and managing conflicts (\cite{luong2011better}). At the same time, challenges arise with modern technology, as older individuals often struggle with digital communication tools. This can lead to feelings of isolation or difficulties in maintaining online social connections (\cite{vaportzis2017older}). Despite these challenges, face-to-face communication remains vital for emotional well-being, as it fosters belonging and reduces loneliness. With age, cognitive and emotional changes can affect communication, for example by slowing response times or altering emotional reactions. Technology can nonetheless offer significant benefits, providing older adults with ways to stay connected when face-to-face communication is no longer feasible (\cite{fuss2019computer, Doering2022b}). Nevertheless, adapting interface design, such as larger text, simplified menus, or adjustable contrast, can substantially increase accessibility and confidence among older users (\cite{choi_internet_2013}). For instance, video calls and social media can help mitigate isolation and support social ties. Ultimately, while older adults may face initial barriers in adapting to new communication forms, their life experience often enriches interactions. With increasing comfort in using technology, they can benefit from both traditional and digital modes of communication (\cite{hulur2020rethinking}). This developmental perspective highlights how age-related characteristics interact with contextual and technological conditions, underscoring the importance of lifespan-sensitive, inclusive communication design in increasingly technology-based communication contexts. Designs that increase the salience of gaze, lip movements, and clear prosody can partially compensate age-related declines, especially in video and VR.\\

Taken together, understanding IC requires a dual focus on psychological and communication-technological perspectives. Psychological perspectives clarify the cognitive, social, and affective demands that communication places on individuals, whereas media-technological perspectives show how these demands are shaped, supported, or disrupted by the properties of communication systems. The interplay between human capacities and technical affordances is especially salient when interaction is mediated by technology, an increasingly common condition across personal, educational, and professional domains. Building on these psychological foundations, we now turn to the media-technology perspective, which specifies the technical conditions that enable, constrain, or transform the shared speech, acoustic, and nonverbal signals through which these communicative processes unfold.

\subsection{A Technology Perspective on IC} \label{TechnologyPerspective}
Media technologies fundamentally shape the conditions and possibilities of IC. They enable new forms of communicating and can serve as substitutes when face-to-face communication is difficult or impossible. They extend and supplement face-to-face interaction by enabling communication across time, space, and modality—shaping the availability, fidelity, and synchrony of communicative signals. From a functional perspective, media technologies are not only transmission channels, but also shaping environments that enable, constrain, or transform communicative practices. They modulate how verbal and nonverbal information is produced, perceived, and integrated, and thereby influence critical aspects such as turn-taking, presence, and mutual understanding. In contrast to Section~\ref{Psychology}, which emphasized psychological mechanisms, the present section highlights media-technological conditions that shape the shared signal substrate of IC., We organize this perspective along the main technical layers through which communication is mediated: audio and acoustics, which determine the preservation and enhancement of speech signals; video, which adds visual access to nonverbal cues; virtual reality, which integrates spatial, embodied, and multimodal interaction; and auditory virtual environments, which provide a focused case for how spatial audio and rendering fidelity shape IC. Across these layers synchronicity, modality, and immersiveness, shape whether media technologies afford or hinder IC across conventional and technologically augmented contexts. To illustrate, we use the following terms: \emph{latency} (end-to-end delay) disrupts turn-taking; \emph{jitter} (delay variability) destabilizes timing; \emph{packet loss} degrades intelligibility; and \emph{audio-visual desynchronisation} misaligns lip and speech cues. Each of these aspects directly impacts coordination in IC.

\subsubsection{Audio and Acoustics} \label{AudioAcoustics}
Audio technologies directly determine how speech and other acoustic signals are preserved, enhanced, or distorted in transmission. Dedicated measures of audio processing and improvements of acoustics can substantially enhance IC. In this context, speech and its intelligibility are crucial. For example, in a large hall with multiple interactive partners, face-to-face communication can be facilitated for interlocutors through the use of microphones and loudspeakers. However, if these tools are not used optimally, e.g., introducing noise, acoustic echo, or audio/audio and audio/visual desynchronization, IC can suffer.

For CMC, audio needs to be converted from an analog to a digital signal, transmitted and then reconverted back to analog form. Through digital signal processing, the output signal can be enhanced to improve speech intelligibility and facilitate IC (\cite{Vary24}). However, if transmission introduces artifacts such as delays or distortions, IC may be impaired and listening fatigue can increase (see, e.g., work reviewed in \cite{Raake2022}). Echo and insufficient acoustic echo cancellation (AEC) further increase effort and disrupt turn-taking. Modern beamforming and adaptive filtering algorithms can mitigate these effects by dynamically emphasizing target speech while suppressing diffuse background noise (\cite{Doclo2015}).

Voice communication via telecommunications has a long tradition, beginning with classical telephony (Plain Old Telephony Service, POTS). While still available today through modern devices such as smartphones, it has been significantly improved. The expansion of speech bandwidth from narrowband (300-3400 Hz) to wideband and further to fullband (up to 7 kHz or 20 kHz), has markedly improved speech quality (e.g., \cite{Raake2007}). This broader bandwidth also enables a more face-to-face-like communication. For example, when spelling names, speech sounds such as /f/ and /s/ can now be more easily distinguished, reducing the need for POTS-style spelling aids such as the NATO alphabet (e.g., Foxtrot for /f/, Sierra for /s/). Improved speech quality can also influence social perception, making interlocutors appear more positively or moderating specific vocal attributes (\cite{Gallardo2018}).

The challenges of attentional selection and inhibition in complex auditory scenes, discussed in Section~\ref{Psychology}, also inform the design of media technologies. Often, several potentially competing acoustic sources are available for auditory processing simultaneously, such as different people speaking at a cocktail party, where the ability to focus on a single target speaker among multiple competing talkers is known as the cocktail party effect (\cite{Cherry1953,pichora2017older}). In this situation, a strong attentional selection of the relevant acoustic information is needed (see \cite{shinn2008object} for a review), while the cognitive processing of irrelevant information needs to be sufficiently suppressed (inhibited). From a technological perspective, this underscores the importance of spatial hearing support (e.g., binaural rendering, advanced room acoustic design) to facilitate selective attention. Here, the interactivity of the auditory scene seems to be particularly important.  The human ability to selectively focus attention on a specific sound source in a complex acoustic scene strongly depends on spatial hearing (\cite{Blauert1996}). Thus, when more acoustically complex scenes are to be considered (e.g. with room acoustics, cf. \cite{Oberem2018a}, moving sound sources or moving listener), the validity of statements on attentional processes may profit significantly.
Speech-based communication between an interactively changing speaker and listener can be viewed as the test case of auditory cognition “in action.” It requires a wide range of functions and processes across human information processing, from speech perception (e.g., in a noisy environment, requiring scene analysis) to cognitive information processing of message plus context information (e.g., affective-emotional connotations) up to response evaluation with respect to the interaction partner (e.g., gender, status). Interactive virtual environments provide a particularly powerful testbed here: by combining complex auditory (e.g., individualized interactive sound presentation) and audio-visual input, they allow realizing communication scenarios of much higher complexity and ecological validity than classical experimental setups. For example, Devesse et al. (\citeyear{devesse.2018}) used virtual humans  to explore the effects of audio-visual cues on speech intelligibility in adults. Similarly, studies with children (\cite{Nirme.2018, Breuer2022, Seitz2024}) examined speech perception and listening effort in noisy classroom-like scenarios, testing whether visual presentation of a virtual speaker could compensate acoustic challenges.

With a focus on IC and age, hearing aids play a crucial role in enabling individuals with hearing impairments to participate more fully or in some cases even at all in IC. This is particularly relevant for older adults, as age-related hearing loss is associated with reduced social participation and has been identified as a contributing factor to cognitive decline, dementia, and other health issues (\cite{pichora2015hearing}, \cite{Livingston2024}, see also \ref{AssistiveListeningDevices}).

\subsubsection{Video}\label{video}
Video enhances IC by providing richer and more dynamic content, thereby supporting understanding and fostering engagement and collaboration. In particular, video signals allow access to visual nonverbal cues, supporting the interpretation of communicative features such as irony or turn-taking signals (\cite{kendrickTurntakingHumanFacetoface2023, aguertParaverbalExpressionVerbal2022}). Reduced frame rates and audio-visual desynchronization have been shown to elongate conversational turns and increase perceived cognitive effort (\cite{framerateJackson}). Empirical studies demonstrate that visual turn-taking cues can reduce listening effort and may even compensate for age-related declines in listening under acoustically challenging (e.g., cocktail-party listening, \cite{pichora2017older}) conditions (\cite{getzmann2017visually}). However, poor synchronization of audio-visual speech signals can create conflicts between modalities (\cite{seurenWhoseTurnIt2021}), which significantly impair speech comprehension, particularly in older adults (\cite{begau2022role}), underscoring the need for reliable technical solutions in multimodal speech presentation. Camera framing and angle modulate perceived eye contact (and thus turn-yielding), while visual clutter increases search load and can mask subtle facial cues.

\subsubsection{Virtual Reality} \label{VR}
Virtual Reality refers to a computer-generated, three-dimensional, and interactive environment that enables users to experience a sense of “being there” (\cite{slaterSeparateRealityUpdate2022}) in a digital space. VR is often mediated through head-mounted displays (HMDs), but projector-based VR systems like CAVE systems (Cave automatic virtual environment), for example with up to six projector sides surrounding the user, are also available. Embodied interaction is a key feature of VR. It is enabled most basically via real-time head tracking, which lets users perceive a virtual environment from their own point of view, but it also extends to hand, body, gesture, and locomotor interaction. This bodily coupling underlies the sense of embodiment, the experience of a virtual body as one's own \parencite{kilteniSenseEmbodimentVirtual2012}. Tracking is used to translate users’ actions (e.g., head movements) to consequences (changes in visual display) in the virtual world (i.e., by establishing  sensorimotor contingency; \cite{oreganSensorimotorAccountVision2001}). VR can thus evoke the psychological state of presence (\cite{slaterPlaceIllusionPlausibility2009}) which includes the illusion of “being there” (place illusion) and the illusion that things around one are actually happening (plausibility illusion; \cite{slaterSeparateRealityUpdate2022}), and has been shown to increase in VR compared to computer presentation (\cite{Ermert2025a}). Presence further depends on real-time eye-tracking and naturalistic gaze behavior, which enhance co-presence and trust between virtual interlocutors (\cite{slaterSeparateRealityUpdate2022, ruboSocialGazeFingerprints2025}). In contrast to traditional media, which do not evoke such illusions, this gives users the ability to perceive and act in ways that closely mirror real-world scenarios. Importantly, while VR research has focused on visual stimulation, other modalities such as auditory (see Section \ref{AuditoryVirtualEnv}), haptic (\cite{fermoselleLetsGetTouch2020}) or olfactory (\cite{tewellReviewOlfactoryDisplay2024}) information are increasingly used to provide rich, multi-sensory experiences. This makes VR a powerful medium for studying and facilitating IC.

VR is especially promising for IC dbecause it can combine speech, spatial audio, gaze, gesture, posture, facial expressions and interpersonal distance within one controllable communication environment. VR allows to implement face-to-face social interactions where both verbal and nonverbal information are integrated. Partly, this resembles the advantage of video communication over audio-only communication resulting from additional information, for instance via facial expressions and gestures. Crucially, however, due to its unique use of spatial cues VR can provide information that goes beyond screen-based video communication. This includes the use of gaze information (e.g., establishing eye contact; \cite{ruboSocialGazeFingerprints2025}), body posture (e.g., leaning toward one) and interpersonal distance (e.g., establishing a comfortable distance to a stranger, i.e., enabling more real-life-like proxemics control) that are central to building rapport and regulating conversational flow (\cite{argyleBodilyCommunication2013}). The fidelity of these cues strongly influences how participants interpret intentions, trustworthiness, and emotional states and has been related to social presence (\cite{ohSystematicReviewSocial2018}). Social presence emphasizes the perception of others as co-present and “real” within that environment (\cite{bioccaMoreRobustTheory2003}). High levels of presence are linked to deeper engagement, stronger emotional responses, and more natural communication patterns (\cite{sprottAvatarsPhygitalSocial2025,pfallerSocialPresenceModerator2021}). Conversely, when presence is weak, communication can feel artificial and disengaging.

The interactive and multimodal nature of VR environments can, however, also introduce challenges to IC. For (virtual) social agents, the phenomenon of the uncanny valley has been described, where agents that are almost, but not perfectly, realistic can evoke discomfort or eeriness (\cite{mori2012uncanny}). Small mismatches in facial expressions, timing of movements, or lip synchronization can disrupt the illusion of authenticity and hinder interaction. This makes the design of avatars and agents a critical factor for social scenarios in VR. Another important aspect in VR-mediated IC concerns timing and synchronization. Delays in speech transmission, gesture rendering, or avatar/agent animation can disrupt conversational interaction \parencite{cortesDelayThresholdSocial2024}.
Achieving low-latency communication systems is therefore essential for creating fluid and effective VR interactions.

In sum, VR offers a unique platform for studying and enhancing IC in complex audio-visual environments (\cite{Rubio-Tamayo2017immersive, Breuer2022, Breuer2025, Ermert2025c, Schiller2024}), which allows the examination of challenging conversational situations in a controlled and reproducible manner. Its potential lies in its ability to replicate and extend social cues and provide controlled environments for experimentation, while also demanding careful design to avoid pitfalls such as the uncanny valley and timing disruptions.

\subsubsection{Auditory Virtual Environments}\label{AuditoryVirtualEnv}
Auditory VR provides a particularly sharp focus on IC, as choices regarding the design of the audio rendering engine can critically influence key aspects such as the localization of speakers, the sense of presence, the management of turns, and the integration of verbal and non-verbal cues (\cite{kotheEffectAvatarHead2025,rosskopfImpactBinauralAuralizations2024}). In the following section, we focus on spatial audio, which has significant implications for these processes.

The central goal of spatial audio in VR is to create a realistic representation of an acoustic scene. This can be achieved through a perceptual illusion whereby virtual sound sources are experienced as if they were real (\cite{blauert_technology_2020}), or through credible perception when there is no real counterpart. This plausibility is not only a technical aim but also highly relevant for IC as it increases presence, supports users in orienting to multiple speakers, following conversational turns, and potentially coordinating joint activities in virtual spaces, much as they would in real-life face-to-face settings (\cite{rungtaEffectsVirtualAcoustics2018,hendrikseMovementGazeBehavior2019}).

In acoustics, there are a variety of tools and methods to create auralizations. Regardless of the setting, the aim is to auralize an acoustic scene (\cite{Vorlaender08}), which can be achieved using measurements or simulations (\cite{wendt2014a, Gergen2012, llorca2021multi,grimm2019toolbox,institute_for_hearing_technology_and_aco_2024_13788752}).

One possible method is to measure impulse responses (IRs) at the listener's ears in order to recreate the underlying sound field and, with this, how a person would encounter sound in this situation. This process requires both a source and a receiver. For illustration, consider the case of a stationary loudspeaker in a room directed towards a human listener and using a specific directivity pattern. Real-world sound sources can add further complexity as they may move and display time-varying directivity, such as a speaking person (\cite{Ehret2020i}).

The receiver may be a real person, with microphones placed at the eardrum or at the entrance of the ear canal to capture what that individual would hear. The measured impulse responses for the left and right ears then represent a specific scenario defined by the exact placement and orientation of both the source and the receiver. To generalize these measurements, head and torso simulators (HATS) are commonly used as standardized representations of human listeners. These devices replicate the acoustic properties of the human head and torso, allowing consistent and repeatable impulse response measurements in different acoustic environments. For IC studies, this standardization makes it possible to test communication scenarios in a reproducible way while still approximating realistic auditory input.

To create more realistic conditions, for example, allowing the receiver to rotate their head, multiple IRs must be measured with different head orientations. Here, the azimuthal resolution is a crucial parameter, which is typically chosen between one to five degrees. Allowing additional degrees of freedom (DoF), such as moving within the room, increases measurement complexity, as every possible alignment and position combination must be captured. At the same time, it allows for a more precise approximation of the characteristics and dynamics of real IC. 

To play back auralizations, these measured responses are convolved in real-time with anechoic stimuli, adapting to both source and receiver position and orientation. If either the source or the receiver moves, the system must switch the IRs in real time, without introducing audible artifacts. An overview of several partitioned convolution algorithms can be found in (\cite{wefers_partitioned_2015}). At least for head movements, the time-variant overlap-add in partitions algorithm  (\cite{jaeger_echtzeitfaehiges_2017, jaeger_timevariant_2023}) can be employed to incorporate dynamic head-tracking during audio playback. Such low-latency adaptation is essential for IC, as delays or artifacts can disrupt conversational flow and the perception of being co-present.

Depending on the number of DoFs needed and the source and receiver combinations, the measurement of IRs may be very time-consuming. One possibility to overcome this is to simulate the acoustics for the targeted scene.

For indoor environments, rooms can be auralized over headphones (e.g. \cite{Masiero2011}) by utilizing human anthropometric characteristics (cf. \cite{Fels2009}) encoded in head-related transfer functions (HRTFs) and incorporating spatial information through binaural room impulse responses (BRIRs) (\cite{xie2013head,blauert_technology_2020}). 
Through this process of auralization, listeners perceive sound sources as if they were actually present in the room (\cite{Oberem2016,brinkmann_assessing_2014,starz_comparison_2025}).

Various methods are available for simulating rooms, each with distinct advantages and limitations. While some approaches focus on physical accuracy, others aim to reduce computational complexity and instead provide perceptually plausible renderings of room acoustics. Consequently, the computational effort and thus the time required to compute IRs, or more specifically BRIRs, can vary substantially.

Some tools are capable of rendering in real-time (\cite{dehaas_realtime_2025,Aspoeck2014,grimm2019toolbox,institute_for_hearing_technology_and_aco_2024_13788752}). For such applications, parameters such as room geometry, absorption coefficients, reverberation time, and the positions of sources and receivers must be specified. The resulting room acoustics can be computed on the fly, enabling immediate auralization with dynamic head-tracking of the listener.

By contrast, if the simulation tool does not support real-time rendering—often due to the higher computational cost or the need for greater physical precision, BRIRs must be precomputed and stored in a dataset. These datasets are then used for playback and auralization, in the same way as measured impulse response datasets. 
Recent advances in Ambisonics and individualized HRTFs further improve spatial accuracy and listener immersion, contributing to more realistic turn-taking and localization cues (\cite{blauert_technology_2020}).
Taken together, auditory VR provides a powerful testbed for studying how technological parameters shape the perceptual and interactional foundations of IC; both perspectives emphasize that successful IC emerges from the alignment between human perceptual capacities and the technical fidelity of the communication channel.

\section{Measures and Context} \label{Measures}
Examining IC from both media-tech\-nological and psychological perspectives provides a comprehensive theoretical foundation for a multidimensional analysis of IC processes. Building on this foundation, the following section outlines key measurement approaches that capture verbal and nonverbal signal quality, communicative synchrony, and experiential correlates such as cognitive effort or perceived presence. Each family targets a different facet of the constructs introduced in Section~\ref{Theory}: signal-quality measures index how well the speech and nonverbal signal is preserved, synchrony measures capture the temporal coupling and turn-taking that coordinate an exchange, and experiential measures track its perceived quality, such as effort, presence, and rapport. Because no single family captures IC on its own, we treat them as complementary and recommend triangulating across them.

\subsection{Behavioral Measures} \label{BehavioralCognitiveMeasures}
Behavioral approaches in the study of IC focus on directly observable actions, whereas cognitive measures aim to infer underlying mental processes. In practice, these domains are closely intertwined, as overt behavior often serves as an indirect indicator of cognitive or affective states.

\subsubsection{Communication Structure and Timing} \label{CommunicationStructureTiming} 
The temporal structure of conversation, who speaks when, for how long, and how turn exchanges are managed, provides fundamental insights into the quality and coordination of an interaction. Analyses are typically based on audio recordings. Initial processing often involved voice activity detection (VAD) algorithms (\cite{sohn1999statistical}) to identify basic speech segments. However, contemporary approaches increasingly leverage sophisticated automatic speech recognition (ASR) and natural language processing (NLP) models, such as those exemplified by OpenAI's Whisper (\cite{radford2023robust}). These advanced tools can provide not only precise VAD but also accurate transcription, speaker diarization (identifying who spoke when), and extract richer hierarchical linguistic features (e.g., word timings, pause durations within and between utterances). This detailed output allows for more nuanced modeling of the interaction as a sequence of conversational states, such as silence, single-speaker monologue, or overlapping speech (\cite{heldner2010pauses,Brady1965}). The frequency and duration of these states vary depending on the interlocutors, the situational context, and the conversational goals. Conversation-analytic studies have long shown that turn-taking follows systematic timing rules that reflect both cognitive prediction and social coordination (\cite{sacks}). Deviations from typical state patterns may indicate communicative strain; prolonged silences, for instance, can be perceived as disengagement, while high levels of overlapping speech might reflect either increased involvement or disruption, depending on the context (\cite{levinson1983pragmatics}). Also, delayed reactions represented by longer pauses than expected may even be related with changes in which personality traits are associated with the other talker (\cite{Schoenenberg2014}). Such timing-based personality inferences highlight how micro-delays in speech contribute to perceived social attributes like warmth, dominance, or competence. Transitions between these states are analytically significant, as they often correspond to moments where coordination may succeed or fail. Key phenomena include within-turn pauses, silent intervals between turns, and interruptions, each of which can vary in appropriateness and interpretation depending on the situation. Commonly derived parameters include the probability of particular states, their mean durations, the frequency of speaker alternations, and the latency between turns. Measures such as the rhythmicity of utterances or the extent of overlapping speech can serve as indicators of conversational fluency and engagement (\cite{couperkuhlen2001interactional}). In experimental contexts, these metrics are sometimes supplemented with manual indicators (e.g., button presses) to mark perceived turn transitions. Perceptual measures such as the Mean Opinion Score (MOS) for conversations, with methodologies discussed in foundational works on spoken dialogue system quality, can further provide subjective assessments of conversational quality (\cite{mollerAssessmentPredictionSpeech2000}). Conversational MOS typically captures perceived smoothness, responsiveness, and overall ease of dialogue in addition to intelligibility (\cite{ITU-T_P800.1_2016, Naderi2024MultiDimensional}).

\subsubsection{Nonverbal and nonvocal Behavior}
Nonverbal/nonvocal behavior is captured via manual annotation or automatic landmark/pose tracking (e.g., OpenPose, MediaPipe) and motion capture. Analyses quantify frequency and timing of cues, kinematics, and interpersonal synchrony using cross-correlation, dynamic time warping, or windowed agreement metrics; end-to-end toolkits (e.g., envisionBOX) support such pipelines. Head movements such as nods or shakes can signal agreement, understanding, or shifts in attention. Gaze plays a central role in establishing joint attention and structuring turn transitions (\cite{kendon1967some}), a concept theoretically grounded in Section~\ref{CommunicationStructureTiming}. Methodologically, gaze can be quantified in terms of fixation patterns, saccade characteristics, pupil diameter, and blink rate, using either remote or head-mounted eye-tracking systems. Beyond individual gaze metrics, interpersonal gaze coupling—temporal alignment of eye movements between partners—has emerged as a robust marker of engagement and mutual understanding (\cite{richardson}). Electrooculography (EOG) offers an alternative method based on the measurement of corneo-retinal potentials, often used in conjunction with EEG to monitor eye movements (\cite{picton2000guidelines}). Interpersonal synchrony is operationalized as temporal alignment between partners’ movement or gaze signals and linked to perceived rapport in validation studies. 

A particular challenge in interpreting nonverbal signals lies in their context dependency and interindividual variability. The same gesture or posture may carry different meanings depending on cultural background, personality traits, or conversational setting. As a result, normative baselines for comparison are difficult to establish, and the communicative relevance of a particular behavior often depends on its temporal and functional embedding within the interaction. Cultural and individual variation should also be considered; for example, the same gesture may signal agreement in one culture but hesitation in another (\cite{argyleBodilyCommunication2013}).

\subsection{Neural Measures}
We structure neural measures from classic event-related potentials (ERPs) derived from the EEG (e.g., N1/P2/N400) to oscillatory indices (e.g., alpha suppression), neural speech tracking and multivariate temporal response functions (mTRF), auditory attention decoding, and hyperscanning.

All these neuroscientific methods provide insight into the brain activity underlying communicative processes and enable the investigation of cognitive mechanisms that are not directly accessible through behavioral observation. Electroencephalography (EEG) is particularly well-suited for studying IC due to its high temporal resolution, which allows for the analysis of fast neural dynamics during processes such as listening, speaking, and turn-taking (\cite{babiloni2014social}). Methodologically, EEG data are typically acquired using electrode caps of varying densities. Recent developments in mobile EEG have expanded the methodological possibilities for investigating communication in ecologically valid settings. In contrast to traditional lab-based EEG systems, mobile EEG enables participants to move freely or engage in naturalistic tasks while brain activity is continuously recorded. This is especially valuable for capturing dynamic, real-world behaviors such as joint attention, gesture use, or turn-taking as they unfold in context. Mobile EEG systems therefore offer practical advantages for naturalistic research but also introduce specific challenges, particularly related to motion artifacts and speech-related muscle activity. Research has shown that meaningful neurophysiological data, including indices of auditory attention and speech tracking, can be reliably obtained even in less controlled environments (\cite{debenerMobileEEG}). For instance, dual-EEG ‘hyperscanning’ has revealed inter-brain synchrony patterns that correlate with cooperation and mutual attention (\cite{dumas, babiloni2014social}). A central focus of current EEG research in communication is neural speech tracking, which refers to the alignment of neural activity with the temporal structure of speech. Slow cortical oscillations, particularly in the delta and theta frequency bands, have been shown to phase-align with the speech envelope, reflecting processes related to auditory attention and speech comprehension (\cite{luo2007phase}). EEG is also used to investigate selective auditory attention in multi-talker situations, often by examining event-related potentials (ERPs) or oscillatory dynamics, such as alpha-band suppression during focused listening (\cite{foxe2011role}).

Further studies have explored neural markers of prediction and anticipation that precede expected linguistic input (\cite{kroczek2021time}) or turn boundaries, which are considered essential for the temporal coordination of dialogue (\cite{arnal2012cortical}). In addition, preparatory motor activity, including readiness potentials, can be observed before speech or gestural contributions, indicating intention to interact. In dyadic settings, EEG has also been employed to study inter-brain synchronisation. Correlated neural activity between individuals is interpreted as a marker of shared attention, alignment, or mutual engagement (\cite{hasson2012brain}). Quantitative measures commonly derived from EEG include ERPs time-locked to communicative events, lateralisation patterns, and spectral power analyses associated with attention and affect. Increasingly, multivariate temporal response function (mTRF) models are used to estimate the mapping between continuous speech features and neural responses, offering a measure of how closely brain activity tracks different aspects of the speech signal (\cite{crosse2016multivariate,daeglau2025}). In hyperscanning paradigms, coherence and correlation metrics are computed between individuals to assess neural synchrony (\cite{babiloni2014social,GrimmOVhyper2024, zammhyper}).

In addition to EEG, functional near-infrared spectroscopy (fNIRS) is gaining popularity in studies of IC. fNIRS measures changes in cortical blood oxygenation and provides higher spatial specificity than EEG, while offering tolerance to movement. It is therefore well-suited for naturalistic and mobile scenarios and is increasingly used in combination with EEG to complement temporal and spatial aspects of neural data (\cite{cui2012nirs}). Like EEG, fNIRS recordings can be affected by superficial physiological noise. In both fNIRS and EEG, precise temporal synchronisation between neural and behavioral signals is essential. When comparing across individuals or sessions, careful preprocessing and normalisation are required to ensure data quality and interpretability.

\subsection{Other Physiological Measures}\label{OtherPhysiologicalMeasures}
In addition to central nervous system activity, physiological signals from the peripheral nervous system can provide valuable information about the internal states of individuals during communication. Such measures are particularly relevant when emotional or embodied aspects of interaction are of interest, and they often complement neural and behavioral data. 
Electromyography (EMG) enables the detection of subtle muscular activity that is not necessarily visible but may reflect preparatory or emotional processes relevant to communication (\cite{fridlund1986guidelines}). For instance, pre-speech motor activation can be observed in facial or articulatory muscles shortly before the onset of vocalisation, reflecting processes of articulatory planning and preparation (\cite{hickok2012computational}). Such activity may indicate motor planning and readiness to engage. Similarly, low-amplitude muscle responses can reflect spontaneous mimicry of the interlocutor’s facial expressions (\cite{hess2013emotional}). These subtle imitative behaviors are thought to play a role in empathy and emotional alignment, and may occur without conscious awareness. EMG recordings are typically obtained from facial muscle groups, such as the zygomaticus major (associated with smiling), the corrugator supercilii (associated with frowning), or perioral muscles involved in speech production (\cite{cacioppo1986electromyographic}). However, EMG signals are susceptible to various sources of noise, including electrical interference and movement artifacts. Reliable measurement therefore requires careful electrode placement, appropriate signal filtering, and the use of baseline corrections to ensure comparability across individuals and conditions. 
Physiological synchrony, correlated changes in autonomic signals between interacting individuals, is increasingly used as a marker of shared affective states or social alignment. Further measures such as heart rate variability (HRV) and electrodermal activity (EDA) are also commonly employed in this context. 

Physiological synchrony, for instance in heart rate or skin conductance, has been proposed as a marker of emotional resonance or mutual attunement (\cite{palumbo2017interpersonal}). Such synchrony is often observed in emotionally intense or well-coordinated interactions, but its interpretation remains context-dependent. Behavioral mimicry, the unintentional imitation of another person’s gestures or expressions, is another frequently studied phenomenon. It has been linked to affiliation and prosocial orientation and may contribute to the smoothness of interaction (\cite{chartrand1999chameleon}). 
Although these signals do not provide direct access to cognitive content, they offer insights into arousal, stress, and emotional engagement, and are particularly useful in naturalistic or prolonged interaction scenarios.

\subsection{Social and Affective Dimensions}
IC is not only a cognitive but also a fundamentally social and emotional process. Beyond the exchange of information, communication shapes interpersonal relationships and is guided by social expectations, emotional expressions, and mutual interpretations. Accordingly, measures of social alignment, empathy, and affective involvement are central to a comprehensive understanding of interaction quality. Beyond information exchange, interaction quality depends on mutual trust, rapport, and empathy—constructs that can be captured via post-interaction questionnaires (e.g., Social Presence Scale, (\cite{bioccaMoreRobustTheory2003}).
Emotional expressions serve not only as signals of internal states but also as social cues that shape the behavior of others. The degree of alignment between interlocutors can be assessed through a range of behavioral and physiological indicators. Linguistic convergence (\cite{giles1991accommodation}), shared rhythm or posture, and mutual gaze patterns are frequently interpreted as signs of rapport and social cohesion (\cite{tickle1990nature}). Particular emphasis has been placed on constructs such as interpersonal trust and rapport, which are central to sustained collaboration and long-term relationship building. Trust ratings often predict willingness to disclose information or cooperate in subsequent tasks, while rapport is linked to perceptions of conversational smoothness and mutual understanding.
In addition to physiological and behavioral markers, subjective rating scales are widely employed. Common instruments include Social Presence scales, trust/rapport measures, NASA-TLX (effort), and the Self-Assessment Manikin (affect); we recommend triangulating these with behavioral and physiological indicators.
Such measures are often collected immediately after an interaction and can be triangulated with objective indicators to capture the experiential quality of communication more comprehensively. These are typically obtained using Likert or semantic differential scales and can be analyzed alongside physiological data. 
Although these dimensions are difficult to operationalize in a standardized way, they are essential for understanding how communicative exchanges are experienced by the participants and how these experiences shape the course of interaction.

\subsection{Physical Property Analysis}
Because IC ultimately unfolds through observable speech, acoustic, and movement signals, the analysis of their physical properties provides a direct link between communicative behavior and the underlying psychological and technological processes. A detailed examination of the physical signals generated during communication can provide objective insights into both the content and the form of interaction. This includes the acoustic properties of speech, reflecting concepts discussed in Section~\ref{AudioAcoustics}, as well as the kinematics of movement, such as gestures and articulatory actions, which relate to the functions of nonverbal cues outlined in Section~\ref{NonverbalModalitySocialCognition} and Section~\ref{IntegratingVerbalNonverbalInformation}. These features are not merely by-products of communication but carry essential information about speaker identity, emotional state, and interactional intent.
Acoustic analyses of the speech signal typically focus on prosodic features such as pitch, intensity, and timing. These parameters are known to reflect cognitive load, emotional engagement, and interactional alignment. The phenomenon of prosodic entrainment, where interlocutors begin to converge on aspects such as speech rate, intonation patterns, or loudness over the course of a conversation, is of particular interest in studies of social coordination (\cite{pardo2006phonetic}). Measures of fluency, such as articulation rate, the frequency and duration of pauses, and the occurrence of disfluencies, are commonly used as indicators of communicative ease or strain. Spectral features of the signal, including formant frequencies or measures of voice quality such as jitter and shimmer, may also be analyzed to capture fine-grained aspects of articulation and expressivity.

These analyses are typically performed using dedicated software such as Praat (\cite{boersma2001praat}) or OpenSMILE (\cite{eyben2010opensmile}), or within custom pipelines in MATLAB or Python. However, the reliability of acoustic measures is strongly dependent on the recording quality and environmental conditions, and prosodic variability can be substantial across speakers and contexts. In addition to vocal features, movement-based signals such as gestures, head movements, or facial expressions offer further information about communicative intent and coordination. The study of kinematic properties involves tracking the position and trajectory of specific articulators or limbs, often using motion capture systems, pose estimation algorithms, or inertial sensors. Of particular relevance is the coordination between different modalities, for instance between gesture and speech, or between gaze and turn-taking cues. Respiratory patterns during interaction are also increasingly studied, as they are closely linked to speech planning, physiological arousal, and turn-taking (\cite{rochetcapellan2014take}). Measurements of respiratory rate, inhalation-exhalation ratio, or chest expansion can be obtained via belts or airflow sensors and integrated with other signal types to better understand the embodied dimension of communication.

Temporal coordination across modalities can also be assessed in terms of synchrony, i.e., the alignment of verbal and nonverbal signals across time. High levels of synchrony, such as simultaneous head nods, matched speech rhythms, or coordinated breathing, are often interpreted as markers of engagement and shared understanding in interactive settings.
Cross-modal synchrony, the temporal alignment between signals from different modalities or from different individuals, has emerged as an important concept in recent work. The degree to which, for example, a gesture temporally aligns with a prosodic accent, or the extent to which the movements of two interlocutors synchronize during a conversation, can be taken as indicators of communicative efficiency or interpersonal rapport. Advanced analysis techniques such as dynamic time warping, cross-correlation, or recurrence quantification are used to capture these temporal relationships (\cite{louwerse2012behavior}). These analyses require precise temporal synchronisation of multiple streams; the Lab Streaming Layer (LSL) is commonly used for this purpose (\cite{kotheLabStreamingLayer2014}). Physical-property analyses thus quantify the signal layer through which coordination, synchrony, and communicative experience emerge. However, the interpretation of these measures depends strongly on the interactional context, a point we address next.

\subsection{Contextual Considerations for Measurement}
The selection, implementation, and interpretation of measures in IC research must always take the specific context of the interaction into account. Communication is shaped not only by the participants and their intentions, but also by the modality through which interaction takes place (as discussed theoretically in Section~\ref{F2FvCMC} regarding face-to-face communication vs. CMC), the broader situational setting, and individual differences in cognitive, social, and sensory capacities, such as age-related factors (see Section~\ref{Age}).
The methodological challenges in CMC arise directly from these psychological differences detailed in Section~\ref{F2FvCMC}. Even in video-based communication, slight delays, limited field of view, or reduced nonverbal bandwidth can alter the dynamics of interaction in ways that affect the interpretation of timing or behavioral alignment. Age-related factors, theoretically outlined in Section~\ref{Age}, may also influence communication behavior and its measurement. These include differences in familiarity with digital media, auditory and visual processing capacity, or conversational pacing. Such factors can modulate both the observable features of interaction and the relevance or sensitivity of particular measures. Moreover, cultural norms and language background can influence conversational timing, gesture use, or preferred modes of turn exchange, further complicating the interpretation of cross-participant or cross-group comparisons.
In ecological settings, dual-task probes and experience sampling (ESM) capture moment-to-moment burdens and context shifts. A careful contextualisation of methodological decisions is therefore essential. Measures that are informative and appropriate in one setting may be inadequate or misleading in another. This holds particularly true for paradigms that attempt to simulate natural conversation under controlled conditions. The degree to which such settings approximate real-world interaction should be critically assessed, and any limitations made explicit when interpreting results. Contextual awareness thus plays a crucial role in ensuring that the conclusions drawn from IC research are both valid and transferable. We recommend transparent reporting of synchronisation quality and exclusion criteria, and, where feasible, preregistration to constrain analytic flexibility.

\section{Technical Solutions and Applications} \label{Applications}

The following applications illustrate how the psychological and technological determinants developed in Section~\ref{Theory}, and the measures surveyed in Section~\ref{Measures}, translate into practice. We deliberately present three prototypical domains that differ in maturity and how they intervene with IC: assistive listening devices (well-established) primarily enhance or restore the acoustic speech signal, conversational agents (rapidly evolving) generate, interpret and coordinate spoken und multimodal dialogue, and social VR (emerging and visionary) embeds IC in spatial, audiovisual and embodied environments. Together, they highlight current achievements while continuously raising new questions about how psychological mechanisms and media technologies must be jointly considered to design effective, inclusive, and human-centered communication environments.

\subsection{Assistive Listening Devices} \label{AssistiveListeningDevices}
Assistive Listening Devices (ALDs) are designed to enhance sound perception, especially  speech perception, and reduce background noise, enabling better IC in face-to-face communication and CMC scenarios. Indeed, a prototypical and notoriously difficult communication scenario in hearing research is known as the {\itshape cocktail party} (\cite{Cherry1953}). Accordingly, the quest for an assistive {\itshape cocktail party processor}, enabling close-to-natural IC in challenging acoustic conditions, has been a research topic for many decades.

ALDs are helpful for both hearing-impaired (HI) and normal-hearing (NH) individuals. Examples of ALDs for HI people include hearing aids and cochlear implants, aiming to compensate the consequences of hearing loss. However, the success of IC can also be improved for NH individuals, e.g., in acoustically challenging situations like meetings, classrooms, and public events. Here, ALDs can facilitate conversations, enhance listening experiences and reduce distractions. 

Modern hearing aids employ a range of methods to enhance IC. Their signal processing features aim to counteract the effect of hearing loss (e.g., with wide dynamic range compression) and to enhance signals with an emphasis on speech (\cite{Kates2008}). A popular approach is the use of spatial filtering of impinging sound signals (\cite{Gannot2017}). This allows a focus on sound arriving from a certain direction, while attenuating sounds from other directions. In conventional hearing aids, these directions are fixed, with sound from the back being attenuated (\cite{Doclo2015}). After directional filtering, the incoming speech signal is then further enhanced via spectro-temporal noise suppression (e.g., \cite{Martin01, Vary24}). The combination of all processing steps has to satisfy strict constraints with respect to latency (typically $<$ 10 ms) and power consumption ($\sim$ 1 mW). A careful joint optimization of the features mentioned above may further improve the listening experience (\cite{Kortlang2017,May2018}). In addition, wireless connectivity in modern hearing aids allows user-adjustable controls and direct audio streaming from remote microphones and devices, facilitating improved CMC.\\
State-of-the-art research in ALD technology focuses on computational acoustic scene analysis (CASA) with source separation (e.g., \cite{Zohourian16,Hassager2017,May2018,Zhang2025}) in combination with methods for auditory attention estimation. CASA aims to provide detailed knowledge about the number and types of sound sources along with their position in relation to the ALD user. Thus, attention estimation algorithms based on user behavior (e.g., head movements and gaze direction) can use this information and audio streams provided by CASA to enhance and attenuate certain signals. Hence, speakers of interest can be enhanced while interfering speakers may be attenuated. This enables improved IC in situations where attention-agnostic speech enhancement does not lead to the desired result (\cite{Doclo2015}).

The latest state-of-the-art hearing aids are able to track multiple speech sources and use a spatial filter to select the one assumed to be most relevant to the listener. These devices make use of recent findings in response timing and turn-taking (\cite{Holler15}) that suggests an average 200 ms response time in adult dyadic conversations. Then, an automatic conversation analysis based on the response times between an utterance of the hearing aid user and the multitude of acoustic signals received allows the spatial selection of the desired acoustic source.

For NH individuals, ALDs can improve IC as well. For CMC scenarios, active noise cancellation (\cite{Liebich18}) is useful to deal with ambient noise and established in many Bluetooth\textsuperscript{\textregistered} (Bluetooth SIG, Inc., Kirkland, WA, USA) headphones and earbuds. Regarding face-to-face communication, audio systems enable clear conversations in acoustically subpar spaces (e.g. conference halls) when fitted to the acoustic conditions of the room, amplifying speech without back coupling and additional reverberation.

\subsection{Conversational Agents} \label{ConversationalAgents}
Conversational agents (or chatbots), i.e. computer-mediated dialogue systems, are now common in human-computer interaction (\cite{clarkWhatMakesGood2019}). While conversational agents have been around since the 1960s (ELIZA, \cite{weizenbaumELIZAComputerProgram1966}), they have become increasingly popular in recent years by the advent of large language models (LLMs). There is also evidence that conversational agents are treated similarly as real humans (computers as social actors; \cite{nassMachinesMindlessnessSocial2000}). Especially when they are designed to act in an emotionally intelligent fashion they are capable of building a trustful relationship with their users.

LLMs are often used as conversational agents (e.g. ChatGPT). In contrast to earlier versions of conversational agents which used simple pattern matching algorithms, LLMs are context-sensitive and can be used to generate individual responses that allow plausible language-based interactions. Studies have demonstrated that verbal output of a LLM is evaluated as empathetic and likable (\cite{ovsyannikova2025}) and that persons are sensitive to the communicative style and personality traits which are prompted in the language model (\cite{kroczekInfluencePersonaConversational2025,lim2025artificial}). Overall, AI-controlled (embodied) conversational agents have great potential to allow for naturalistic IC with a computer.

Embodied conversational agents combine a computer-mediated dialogue system with an embodied agent, that is a virtual agent or a robot. For instance, in VR one can sit face-to-face  with a virtual person and the output of the dialogue-system is presented as synthesized voice, lip-synced by the agent (\cite{cassell2001embodied}). Importantly, this approach allows including nonverbal information in conversations with the agents (\cite{ehretWhosNextIntegrating2023}). Conversational agents involve the integration of speech recognition, natural language generation, speech synthesis, animation, facial expression, gaze control, body movement, and spatial audio (\cite{gratch2002creating}). Each of these layers contributes to the user’s experience of agency and realism. For example, a technically accurate language model may still feel socially inappropriate if its voice, facial expression, timing, or gaze behavior does not match the conversational situation. Conversely, even relatively simple dialogue systems can appear socially engaging when embedded in a convincing virtual body and responsive environment \cite{bickmore2004social}. 

Recent research highlights the potential of conversational agents to assist older adults, supporting various aspects of cognitive and emotional well-being. Companion robots using LLMs can, for example, provide social and emotional support, with older adults expecting active engagement, personalization, privacy protection, and empathy (\cite{irfan2024recommendations}), while Talk2Care, an LLM-based voice assistant, can facilitate communication between healthcare providers and older adults (\cite{yang2024talk2care}). Thus, AI-based conversational agents have shown promise in enhancing mental well-being for older adults. However, challenges remain, including the need for increased technological literacy and accessible designs tailored for older adults (\cite{chen2024scoping}). 

\subsection{Social VR}

Many of the above considerations are relevant in the design of collaborative virtual spaces. Social VR aims at closely mirroring real-life social interactions in a virtual environment where users are represented by avatars and where they can communicate in real-time via verbal and nonverbal behaviors (\cite{hennig-thurauSocialInteractionsMetaverse2023}). These properties make social VR a direct application for IC as it allows to engage in direct eye contact or present gestures when communicating with others (\cite{cummingsDistinguishingSocialVirtual2024}). The metaverse, for instance, is a collective virtual shared space created through the convergence of VR and the Internet, and it implements social VR as a key feature. Envisioned as a persistent, immersive, and interactive digital space, it provides an environment where people can work, play, and socialize using embodied conversational agents and other digital assets. Typically, users interact in real-time using avatars, voice, text, and gestures, ideally feeling a sense of presence with others (\cite{sonSocialVirtualReality2025}). In the metaverse, users can also create, buy, sell, and own digital goods, individually designed spaces, and share experiences. Preferably, it is used via a VR headset but may also be accessed through a desktop computer display and an audio headset. While the term metaverse has been coined in Neal Stephenson's 1992 science fiction novel {\itshape Snow Crash}, many commercial implementations of these concepts are also available, for example, (\cite{secondlife}, \cite{meta-metaverse}, \cite{microsoftmesh}).  Although the metaverse aims at creating and supporting a digital economy, also using cryptocurrencies and virtual goods, the evolution of its economic or social value is difficult to predict at this time. However, it already has specific applications, for instance, in training technical personnel for comprehensive assembly and service processes, also in conjunction with AR technologies (\cite{teamviewer}).

While there is accumulating evidence that social VR provides users with interactive experiences that match real-life encounters, it is still a matter of debate how such virtual interactions can be optimally presented while ensuring safety and privacy (\cite{WhitePaperIEEE2021}, see also Section \ref{Ethics}). These challenges need to be addressed when social VR is used for IC. Interestingly, previous attempts to characterize IC in social VR have largely focused on visual aspects of interactive cues like gaze or avatar appearance, while acoustic features have been mostly neglected. Especially, virtual acoustics may play a crucial role here, as real-time reproduction of room acoustics may add context information on communicative situations (e.g.,  being in an office or the terrace of a cafe) and spatial audio rendering can give additional cues on who is speaking in a multi-talker scenario (see Section \ref{AuditoryVirtualEnv}).  

Taken together, social VR and its implementation in the metaverse are likely to have a major impact on IC scenarios in the future. However, additional research is required to specify how general principles of IC translate to social VR and which policies are required to ensure safe interactions.

\section{Ethical Considerations} \label{Ethics}

As IC technologies increasingly shape how people engage with one another (across domains like healthcare, education, or customer service) they raise pressing ethical concerns. These systems do not merely transmit content; they co-construct interaction, often invisibly. A special role is played by VR, which makes it easy to capture a wide range of complex behaviors, such as body movements, gaze, or gestures, on a scale not possible in traditional settings. While this offers rich opportunities for analysis and system improvement, it also amplifies ethical concerns about data use, surveillance, and user autonomy. This makes it essential to reflect critically on how IC systems affect user autonomy, privacy, inclusion, and social trust, especially when working with populations that are already marginalized or more susceptible to harm.

A core ethical concern in both the use and study of IC systems is the handling of sensitive, multimodal data. Interactive systems typically rely on input from multiple channels like speech, gaze, facial expression, movement, physiological responses, and this data can reveal more than users intend. Accordingly, informed consent procedures must be more than an administrative formality. For populations discussed in  Section \ref{AssistiveListeningDevices} for hearing impairments and in Section \ref{ConversationalAgents} for adaptive agents (e.g., older adults with declining cognitive capacities or hearing impairments), standard consent protocols may not suffice. Layered or adaptive explanations, repeated consent checkpoints, and easily accessible privacy settings should be part of ethical design and data collection (\cite{nissenbaum2011contextual}).

Researchers and developers alike should adopt a data minimization approach and favor local or on-device processing where feasible. Data handling policies must align not only with legal standards such as the General Data Protection Regulation (GDPR, \cite{voigt2017eu}), but also with user expectations and lived experiences. In fact, given the uneven global landscape of privacy regulation, ethical practices in IC research must go beyond compliance, ensuring transparency and protection even where no strict legal requirement exists. When AI models are involved, transparency-by-design is essential. This becomes especially important for the kinds of predictive or adaptive systems (such as conversational agents) discussed in Section~\ref{ConversationalAgents}, where the system may modulate turn-taking or adapt to user cues in real-time. Without transparency, these mechanisms risk becoming invisible forms of behavioral steering (\cite{lundberg2017unified,gratch2007can}).

Moreover, the capacity of IC systems to analyze, predict, or even guide user behavior raises ethical concerns around manipulation and agency. For instance, conversational agents that adaptively mirror emotional tone or anticipate responses—discussed in Section~\ref{ConversationalAgents}—may blur the line between support and subtle coercion (\cite{bailenson2018experience}). This is particularly relevant when studying groups with reduced assertiveness or increased susceptibility to social pressure, such as children or users with cognitive vulnerabilities. Researchers should therefore consider embedding real-time transparency indicators, and evaluating systems not just on efficiency but on perceived autonomy and trust.

Bias is another major issue—both in system performance and in the assumptions embedded in communication norms. As shown in Section~\ref{AssistiveListeningDevices}, certain groups, such as neurodivergent individuals, non-native speakers, or those with atypical prosody, are particularly at risk of being misrepresented or misinterpreted by IC systems trained on normative data (\cite{koenecke2020racial,hovy2016social}). These biases are not always evident during development but can surface subtly in interaction, leading to exclusion, discomfort, or reduced usability. We strongly recommend incorporating systematic bias audits and cross-group validation as part of any IC research or deployment pipeline.

The ethical imperative extends beyond the individual to questions of inclusivity and access. As emphasized in Section~\ref{ConversationalAgents}, IC systems should accommodate diverse communication needs, including those of individuals with hearing loss, speech differences, neurodivergence, and culturally divergent turn-taking norms. Usability testing must include these groups from the outset—not as afterthoughts. Furthermore, assistive technologies must be designed with an eye toward empowerment over dependency, avoiding surveillance-style monitoring or extractive data practices (\cite{dignum2019responsible}).

To ensure ethical relevance and social value, IC research should also adopt participatory design approaches wherever feasible. This means engaging users, not just as participants, but as co-designers, particularly those who are frequently excluded from technology development processes. These methods are especially valuable when working with neurodivergent users, older adults, or individuals with limited digital literacy, whose needs are often underrepresented in mainstream datasets and design choices.

In addition, ethical reflection should include the long-term use and impact of IC systems. Many technologies are not deployed in a one-off setting, but rather evolve through updates, model re-training, or ongoing interaction. This requires longitudinal accountability—such as regular bias monitoring, impact assessments, and avenues for user feedback long after initial deployment. Questions of who maintains responsibility for these systems over time—and how users can meaningfully withdraw consent—should be addressed early on.

Finally, as IC systems become more complex and resource-intensive (e.g., multimodal VR setups or LLMs for real-time communication), ethical reflection should briefly consider ecological sustainability and infrastructural justice. Researchers and institutions should remain aware of the energy consumption, hardware dependencies, and digital inequalities that can result from high-tech implementations, particularly in under-resourced settings.

From a research methodology perspective, this means researchers must take responsibility not only for the aspects of IC that can directly be measured, but also for those that remain invisible to their chosen methods. For example, methods such as eye tracking or EEG can provide highly temporally precise data points, but these often reflect only a narrow part of the interaction and may overlook the broader communicative context. A participant might show strong neural markers of listening effort, yet at the same time report feeling socially excluded because their camera was turned of, something that the physiological data alone would not reveal. In line with the reflexive stance discussed in Section~\ref{OtherPhysiologicalMeasures}, we therefore recommend combining quantitative measures with qualitative approaches (e.g., participant interviews, usability diaries) to reveal less visible aspects of IC, such as hidden burdens, unmet needs, or negative side effects. Ethical reporting also requires addressing ambivalence, uncertainty, or emotional responses observed during studies, especially when working with vulnerable populations (\cite{nosek2015promoting}).

Ultimately, ethical reflection in IC research must be as dynamic as the systems it studies. It is a continuous, relational process, not a checklist. The goal is not only to avoid harm but to actively design technologies that support equitable, transparent, and meaningful human connection. This requires shared responsibility across design, deployment, and governance—anchored in accountability, inclusivity, and care.

\section{Summary and Future Directions}

This primer set out to bring together two research traditions that usually work apart: the psychological study of how people coordinate in conversation, and the acoustic and media-technological study of how communication channels preserve or degrade the signals this coordination relies on. Across the definition, theory, measures, and applications, the same theme recurred: successful IC depends on the alignment between human perceptual, cognitive, and social capacities and the fidelity and timing of the channel, with speech and acoustic signal as the place where the two meet most directly. The four criteria included in the definition gave a common reference point, the measurement families of signal quality, synchrony, and experience showed how each facet can be assessed, and the three application domains illustrated how this balance shifts from well-established assistive listening devices to emerging social VR. What still falls between the disciplines is a shared account of how much signal degradation a given coordination mechanism can tolerate before interaction breaks down, and how this depends on the partners, the task, and the medium.

The future of IC will be shaped by the convergence of advanced communication technologies, evolving user needs, and growing ethical awareness, spanning continuing advances in speech and audio processing as well as immersive and AI-mediated platforms. Immersive platforms, such as virtual and augmented reality, promise to bridge the gap between face-to-face and computer-mediated communication by incorporating rich, multimodal signals into digital interactions. However, simulating the natural timing, context sensitivity, and emotional depth of real-life communication remains challenging. AI, especially in the form of LLMs and embodied conversational agents, is transforming the way we interact with machines. These systems are becoming increasingly adept at interpreting and responding to verbal and nonverbal cues, raising questions about trust, empathy, and user autonomy. 

At the same time, research must continue to improve our ability to measure interaction quality, synchrony, and engagement. While advances in mobile methods, neural monitoring, and behavioral analytics open up new possibilities, they also require careful interpretation and ethical oversight. Ensuring inclusivity, privacy, adaptability, and transparency will be as crucial as technical innovation here. Participatory design and interdisciplinary collaboration are key to developing IC systems that are effective, equitable, and respectful of diverse users and contexts.

Finally, it will be challenging to align the rapidly growing technical capabilities of these communication technologies with human communication requirements, considering the individual needs of communication partners and ethical considerations. Close cooperation between different disciplines is required to develop these technical possibilities and evaluate them in terms of user needs. This cross-disciplinary primer highlighted some of these methods, such as modern assistive listening devices and neuro-cognitive methods. The future of IC lies in creating technologies that support genuinely human-centered, adaptive and ethically grounded interaction across both physical and virtual spaces.

\section{Funding Acknowledgement}
This research was funded by the Deutsche Forschungsgemeinschaft (DFG, German Research Foundation) under the project IDs 444832396, 444777670, 444761144, 444831328, 444697733, 444724862, 444532506, 422686707 within the priority program "SPP2236 AUDICTIVE". In addition MD was funded by Cluster of Excellence “Hearing4all" (DFG, project ID 390895286) and AR was funded by the Carl-Zeiss-Stiftung project "CO-HUMANICS".

\section{Conflicts of Interest}
There are no conflicts of interest.

\section{Author Contributions}
Mareike Daeglau: Conceptualization, Project Administration, Writing - original draft, Writing - review \& editing, Visualization. Stephan Getzmann: Conceptualization, Writing - original draft, Writing - review \& editing. Moritz Bender: Conceptualization, Writing - original draft, Writing - review \& editing. Janina Fels: Funding acquisition, Writing - review \& editing. Rainer Martin: Conceptualization, Writing - original draft, Writing - review \& editing. Alexander Raake: Conceptualization, Funding acquisition, Writing - review \& editing. Isabel S. Schiller: Writing - review \& editing. Sabine J. Schlittmeier: Funding acquisition, Writing - review \& editing. Katrin Schoenenberg: Writing - review \& editing. Felix Stärz: Conceptualization, Writing - original draft, Writing - review \& editing. Leon O.H. Kroczek: Conceptualization, Project Administration, Writing - original draft, Writing - review \& editing.

\printbibliography

@article{kilteniSenseEmbodimentVirtual2012,
  author  = {Kilteni, Konstantina and Groten, Raphaela and Slater, Mel},
  title   = {The Sense of Embodiment in Virtual Reality},
  journal = {Presence: Teleoperators and Virtual Environments},
  volume  = {21},
  number  = {4},
  pages   = {373--387},
  year    = {2012},
  doi     = {10.1162/PRES_a_00124}
}

@article{lim2025artificial,
  title={Artificial social influence via human-embodied {AI} agent interaction in immersive virtual reality ({VR}): {Effects} of similarity-matching during health conversations},
  author={Lim, Sue and Schm{\"a}lzle, Ralf and Bente, Gary},
  journal={Computers in Human Behavior: Artificial Humans},
  volume={5},
  pages={100172},
  year={2025},
  publisher={Elsevier}
}

@incollection{bickmore2004social,
  title={Social dialogue with embodied conversational agents},
  author={Bickmore, Timothy and Cassell, Justine},
  booktitle={Advances in Natural Multimodal Dialogue Systems},
  pages={23--54},
  publisher={Springer},
  year={2005}
}

@online{kotheLabStreamingLayer2014,
  title = {Lab {{Streaming Layer}} ({{LSL}})},
  author = {Kothe, Christian},
  date = {2014}
}

@article{gratch2002creating,
  title={Creating interactive virtual humans: Some assembly required},
  author={Gratch, Jonathan and Rickel, Jeff and Andr{\'e}, Elisabeth and Cassell, Justine and Petajan, Eric and Badler, Norman},
  journal={IEEE Intelligent Systems},
  volume={17},
  number={4},
  pages={54--63},
  year={2002}
}

@article{cassell2001embodied,
  title={Embodied conversational agents: Representation and intelligence in user interfaces},
  author={Cassell, Justine},
  journal={AI Magazine},
  volume={22},
  number={4},
  pages={67--83},
  year={2001}
}

@article{dumas,
author = {Dumas, Guillaume and Nadel, Jacqueline and Soussignan, Robert and Martinerie, Jacques and Garnero, Line},
year = {2010},
month = {08},
pages = {e12166},
title = {Inter-Brain Synchronization during Social Interaction},
volume = {5},
journal = {PloS one},
doi = {10.1371/journal.pone.0012166}
}

@article{richardson,
  author  = {Richardson, Daniel C. and Dale, Rick},
  title   = {Looking to understand: The coupling between speakers' and listeners' eye movements and its relationship to discourse comprehension},
  journal = {Cognitive Science},
  year    = {2005},
  volume  = {29},
  number  = {6},
  pages   = {1045--1060},
  doi     = {10.1207/s15516709cog0000_29}
}

@inproceedings{framerateJackson,
  author    = {Jackson, Matthew and Anderson, Anne and McEwan, Rachel and Mullin, Jim},
  title     = {Impact of video frame rate on communicative behavior in two- and four-party groups},
  booktitle = {Proceedings of the ACM Conference on Computer Supported Cooperative Work},
  year      = {2000},
  pages     = {11--20},
  doi       = {10.1145/358916.358945},
  address   = {Philadelphia, PA},
  publisher = {ACM}
}

@article{choi_internet_2013,
	title = {Internet {Use} {Among} {Older} {Adults}: {Association} {With} {Health} {Needs}, {Psychological} {Capital}, and {Social} {Capital}},
	volume = {15},
	issn = {14388871},
	url = {http://www.ncbi.nlm.nih.gov/pubmed/23681083},
	doi = {10.2196/jmir.2333},
	number = {5},
	journal = {J Med Internet Res},
	author = {Choi, Namkee G and DiNitto, Diana M},
	month = may,
	year = {2013},
	pages = {e97},
}

@article{dennis,
author = {Dennis, Alan R. and Fuller, Robert M. and Valacich, Joseph S.},
title = {Media, tasks, and communication processes: a theory of media synchronicity},
year = {2008},
issue_date = {September 2008},
publisher = {Society for Information Management and The Management Information Systems Research Center},
address = {USA},
volume = {32},
number = {3},
issn = {0276-7783},
journal = {MIS Q.},
month = sep,
pages = {575–600},
numpages = {26}
}

@article{sacks,
author = {Sacks, Harvey and Schegloff, Emanuel and Jefferson, Gail},
year = {1974},
month = {12},
pages = {696-735},
title = {A Simple Systematic for the Organisation of Turn Taking in Conversation},
volume = {50},
journal = {Language},
doi = {10.2307/412243}
}

@article{reeves1996media,
  title={The media equation: How people treat computers, television, and new media like real people},
  author={Reeves, Byron and Nass, Clifford},
  journal={Cambridge, UK},
  volume={10},
  number={10},
  pages={19--36},
  year={1996}
}

@article{Kleinke1986,
  title = {Gaze and Eye Contact: {A} Research Review.},
  shorttitle = {Gaze and Eye Contact},
  author = {Kleinke, Chris L.},
  date = {1986},
  journaltitle = {Psychological Bulletin},
  volume = {100},
  number = {1},
  pages = {78--100},
  issn = {1939-1455, 0033-2909},
  doi = {10.1037/0033-2909.100.1.78},
  url = {https://doi.apa.org/doi/10.1037/0033-2909.100.1.78},
  urldate = {2026-07-06},
  langid = {english}
}

@article{Pickering_Garrod_2004, 
title={Toward a mechanistic psychology of dialogue}, 
volume={27}, 
DOI={10.1017/S0140525X04000056}, 
number={2}, 
journal={Behavioral and Brain Sciences}, 
author={Pickering, Martin J. and Garrod, Simon}, 
year={2004}, pages={169–190}}

@book{Clark1996, 
place={Cambridge}, 
series={“Using” Linguistic Books}, 
title={Using Language}, 
publisher={Cambridge University Press}, 
author={Clark, Herbert H.}, 
year={1996}, 
collection={“Using” Linguistic Books}
}

@article{Bailenson2021Nonverbal,
  author  = {Bailenson, J. N.},
  title   = {Nonverbal overload: A theoretical argument for the causes of Zoom fatigue},
  journal = {Technology, Mind, and Behavior},
  year    = {2021},
  volume  = {2},
  number  = {1},
  pages   = {1--6},
  doi     = {10.1037/tmb0000030}
}

@article{FAUVILLE2021100119,
  author  = {Fauville, G. and Luo, M. and Queiroz, A. C. M. and Bailenson, J. N. and Hancock, J.},
  title   = {Zoom exhaustion \& fatigue scale},
  journal = {Computers in Human Behavior Reports},
  year    = {2021},
  volume  = {4},
  pages   = {100119},
  doi     = {10.1016/j.chbr.2021.100119}
}

@book{PsychTelecomm,
author = "Short, John and Williams, Ederyn and Christie, Bruce",
title = "The social psychology of telecommunications",
year = "1976",
publisher = "Wiley",
isbn = "0-471-01581-4",
}

@article{DaftLengel,
author = {Daft, Richard L. and Lengel, Robert H.},
title = {Organizational Information Requirements, Media Richness and Structural Design},
journal = {Management Science},
volume = {32},
number = {5},
pages = {554-571},
year = {1986},
doi = {10.1287/mnsc.32.5.554},
URL = {  
        https://doi.org/10.1287/mnsc.32.5.554
},
eprint = {  
        https://doi.org/10.1287/mnsc.32.5.554
}}

@article{debenerMobileEEG,
author = {Debener, Stefan and Minow, Falk and Emkes, Reiner and Gandras, Katharina and de Vos, Maarten},
title = {How about taking a low-cost, small, and wireless {EEG} for a walk?},
journal = {Psychophysiology},
volume = {49},
number = {11},
pages = {1617-1621},
doi = {https://doi.org/10.1111/j.1469-8986.2012.01471.x},
url = {https://onlinelibrary.wiley.com/doi/abs/10.1111/j.1469-8986.2012.01471.x},
eprint = {https://onlinelibrary.wiley.com/doi/pdf/10.1111/j.1469-8986.2012.01471.x},
year = {2012}
}

@ARTICLE{zammhyper,
  
AUTHOR={Zamm, Anna  and Palmer, Caroline  and Bauer, Anna-Katharina R.  and Bleichner, Martin G.  and Demos, Alexander P.  and Debener, Stefan },
         
TITLE={Behavioral and Neural Dynamics of Interpersonal Synchrony Between Performing Musicians: A Wireless {EEG} Hyperscanning Study},
        
JOURNAL={Frontiers in Human Neuroscience},
        
VOLUME={Volume 15 - 2021},

YEAR={2021},

URL={https://www.frontiersin.org/journals/human-neuroscience/articles/10.3389/fnhum.2021.717810},

DOI={10.3389/fnhum.2021.717810},

ISSN={1662-5161}}

@ARTICLE{daeglau2025,
  
AUTHOR={Daeglau, Mareike  and Otten, Juergen  and Grimm, Giso  and Mirkovic, Bojana  and Hohmann, Volker  and Debener, Stefan },
         
TITLE={Neural speech tracking in a virtual acoustic environment: audio-visual benefit for unscripted continuous speech},
        
JOURNAL={Frontiers in Human Neuroscience},
        
VOLUME={Volume 19 - 2025},

YEAR={2025},

URL={https://www.frontiersin.org/journals/human-neuroscience/articles/10.3389/fnhum.2025.1560558},

DOI={10.3389/fnhum.2025.1560558},

ISSN={1662-5161}}

@INPROCEEDINGS{GrimmOVhyper2024,
  author={Grimm, Giso and Daeglau, Mareike and Hohmann, Volker and Debener, Stefan},
  booktitle={2024 IEEE 5th International Symposium on the Internet of Sounds (IS2)}, 
  title={{EEG} Hyperscanning in the Internet of Sounds: Low-Delay Real-Time Multi-Modal Transmission Using the {OVBOX}}, 
  year={2024},
  volume={},
  number={},
  pages={1-8},
  doi={10.1109/IS262782.2024.10704205}}

@inproceedings{Naderi2024MultiDimensional,
  author       = {Naderi, Babak and Cutler, Ross and Ristea, Nicolae-Cătălin},
  title        = {Multi-Dimensional Speech Quality Assessment in Crowdsourcing},
  booktitle    = {Proc.\ IEEE Int.\ Conf.\ Acoustics, Speech and Signal Processing (ICASSP) 2024},
  pages        = {696--700},
  year         = {2024},
  note         = {Preprint version: arXiv:2309.07385},
  url          = {https://arxiv.org/abs/2309.07385}
}

@techreport{ITU-T_P800.1_2016,
  author       = {{ITU-T Study Group 12}},
  title        = {Mean Opinion Score ({MOS)} terminology},
  institution  = {International Telecommunication Union},
  type         = {Recommendation},
  number       = {P.800.1},
  month        = jul,
  year         = 2016,
  address      = {Geneva, Switzerland},
  note         = {“MOS-CQ = Mean Opinion Score for conversati­onal quality”}
}

@article{hendrikseMovementGazeBehavior2019,
  title = {Movement and {{Gaze Behavior}} in {{Virtual Audiovisual Listening Environments Resembling Everyday Life}}},
  author = {Hendrikse, Maartje M. E. and Llorach, Gerard and Hohmann, Volker and Grimm, Giso},
  date = {2019-01},
  journaltitle = {Trends in Hearing},
  shortjournal = {Trends in Hearing},
  volume = {23},
  pages = {2331216519872362},
  issn = {2331-2165, 2331-2165},
  doi = {10.1177/2331216519872362},
  url = {https://journals.sagepub.com/doi/10.1177/2331216519872362},
  urldate = {2025-11-05},
  langid = {english}
}

@inproceedings{rungtaEffectsVirtualAcoustics2018,
  title = {Effects of Virtual Acoustics on Target-Word Identification Performance in Multi-Talker Environments},
  booktitle = {Proceedings of the 15th {{ACM Symposium}} on {{Applied Perception}}},
  author = {Rungta, Atul and Rewkowski, Nicholas and Schissler, Carl and Robinson, Philip and Mehra, Ravish and Manocha, Dinesh},
  date = {2018-08-10},
  pages = {1--8},
  publisher = {ACM},
  location = {Vancouver British Columbia Canada},
  doi = {10.1145/3225153.3225166},
  url = {https://dl.acm.org/doi/10.1145/3225153.3225166},
  urldate = {2025-11-05},
  eventtitle = {{{SAP}} '18: {{ACM Symposium}} on {{Applied Perception}} 2018},
  isbn = {978-1-4503-5894-1},
  langid = {english}
}

@article{rosskopfImpactBinauralAuralizations2024,
  title = {The Impact of Binaural Auralizations on Sound Source Localization and Social Presence in Audiovisual Virtual Reality: Converging Evidence from Placement and Eye-Tracking Paradigms},
  shorttitle = {The Impact of Binaural Auralizations on Sound Source Localization and Social Presence in Audiovisual Virtual Reality},
  author = {Roßkopf, Sarah and Kroczek, Leon O.H. and Stärz, Felix and Blau, Matthias and Van De Par, Steven and Mühlberger, Andreas},
  date = {2024},
  journaltitle = {Acta Acustica},
  shortjournal = {Acta Acust.},
  volume = {8},
  pages = {72},
  issn = {2681-4617},
  doi = {10.1051/aacus/2024064},
  url = {https://acta-acustica.edpsciences.org/10.1051/aacus/2024064},
  urldate = {2025-02-11},
  langid = {english}
}

@online{kotheEffectAvatarHead2025,
  title = {Effect of {{Avatar Head Movement}} on {{Communication Behaviour}}, {{Experience}} of {{Presence}} and {{Conversation Success}} in {{Triadic Conversations}}},
  author = {Kothe, Angelika and Hohmann, Volker and Grimm, Giso},
  date = {2025},
  doi = {10.48550/ARXIV.2504.20844},
  url = {https://arxiv.org/abs/2504.20844},
  urldate = {2025-11-05},
  pubstate = {prepublished},
  version = {1}
}

@article{cummingsDistinguishingSocialVirtual2024,
  title = {Distinguishing Social Virtual Reality: {{Comparing}} Communication Channels across Perceived Social Affordances, Privacy, and Trust},
  shorttitle = {Distinguishing Social Virtual Reality},
  author = {Cummings, James J. and Shore Ingber, Alexis},
  date = {2024-12},
  journaltitle = {Computers in Human Behavior},
  shortjournal = {Computers in Human Behavior},
  volume = {161},
  pages = {108427},
  issn = {07475632},
  doi = {10.1016/j.chb.2024.108427},
  url = {https://linkinghub.elsevier.com/retrieve/pii/S0747563224002954},
  urldate = {2025-10-24},
  langid = {english}
}

@article{hennig-thurauSocialInteractionsMetaverse2023,
  title = {Social Interactions in the Metaverse: {{Framework}}, Initial Evidence, and Research Roadmap},
  shorttitle = {Social Interactions in the Metaverse},
  author = {Hennig-Thurau, Thorsten and Aliman, Dorothea N. and Herting, Alina M. and Cziehso, Gerrit P. and Linder, Marc and Kübler, Raoul V.},
  date = {2023-07},
  journaltitle = {Journal of the Academy of Marketing Science},
  shortjournal = {J. of the Acad. Mark. Sci.},
  volume = {51},
  number = {4},
  pages = {889--913},
  issn = {0092-0703, 1552-7824},
  doi = {10.1007/s11747-022-00908-0},
  url = {https://link.springer.com/10.1007/s11747-022-00908-0},
  urldate = {2025-10-24},
  langid = {english}
}

@ARTICLE{WhitePaperIEEE2021,
  author={Mark McGill},
  journal={White Paper}, 
  title={{The IEEE Global Initiative on Ethics of Extended Reality (XR) Report} -- {Extended} Reality ({XR}) and the Erosion of Anonymity and Privacy}, 
  year={2021},
  volume={},
  number={},
  pages={1-24},
  doi={}}

@article{ruboSocialGazeFingerprints2025,
  title = {Social Gaze Fingerprints: Identifying Social Virtual Reality Users by Their Eye Gaze Patterns},
  shorttitle = {Social Gaze Fingerprints},
  author = {Rubo, Marius and Son, Gayoung},
  date = {2025-09-02},
  journaltitle = {Virtual Reality},
  shortjournal = {Virtual Reality},
  volume = {29},
  number = {3},
  pages = {144},
  issn = {1434-9957},
  doi = {10.1007/s10055-025-01210-4},
  url = {https://link.springer.com/10.1007/s10055-025-01210-4},
  urldate = {2025-10-01},
  langid = {english}
}

@article{mcgurkHearingLipsSeeing1976,
  title = {Hearing Lips and Seeing Voices},
  author = {McGurk, Harry and MacDonald, John},
  date = {1976-12},
  journaltitle = {Nature},
  shortjournal = {Nature},
  volume = {264},
  number = {5588},
  pages = {746--748},
  issn = {0028-0836, 1476-4687},
  doi = {10.1038/264746a0},
  url = {https://www.nature.com/articles/264746a0},
  urldate = {2025-10-01},
  langid = {english}
}

@article{lombard1911,
  title={Le signe de televation de la voix},
  author={Lombard, Etienne},
  journal={Annu. maladies oreille larynx nez pharynx},
  volume={27},
  pages={101--119},
  year={1911}
}

@article{lane1971,
  title={The {Lombard} sign and the role of hearing in speech},
  author={Lane, Harlan and Tranel, Bernard},
  journal={Journal of speech and hearing research},
  volume={14},
  number={4},
  pages={677--709},
  year={1971},
  publisher={American Speech-Language-Hearing Association}
}

@article{weizenbaumELIZAComputerProgram1966,
  title = {{{ELIZA}}—a Computer Program for the Study of Natural Language Communication between Man and Machine},
  author = {Weizenbaum, Joseph},
  date = {1966-01},
  journaltitle = {Communications of the ACM},
  shortjournal = {Commun. ACM},
  volume = {9},
  number = {1},
  pages = {36--45},
  issn = {0001-0782, 1557-7317},
  doi = {10.1145/365153.365168},
  url = {https://dl.acm.org/doi/10.1145/365153.365168},
  urldate = {2025-09-19},
  langid = {english}
}

@inproceedings{clarkWhatMakesGood2019,
  title = {What {{Makes}} a {{Good Conversation}}?: {{Challenges}} in {{Designing Truly Conversational Agents}}},
  shorttitle = {What {{Makes}} a {{Good Conversation}}?},
  booktitle = {Proceedings of the 2019 {{CHI Conference}} on {{Human Factors}} in {{Computing Systems}}},
  author = {Clark, Leigh and Pantidi, Nadia and Cooney, Orla and Doyle, Philip and Garaialde, Diego and Edwards, Justin and Spillane, Brendan and Gilmartin, Emer and Murad, Christine and Munteanu, Cosmin and Wade, Vincent and Cowan, Benjamin R.},
  year = {2019},
  month = may,
  pages = {1--12},
  publisher = {ACM},
  address = {Glasgow Scotland Uk},
  doi = {10.1145/3290605.3300705},
  urldate = {2025-09-19},
  isbn = {978-1-4503-5970-2},
  langid = {english}
}

@article{mori2012uncanny,
  title={The uncanny valley [from the field]},
  author={Mori, Masahiro and MacDorman, Karl F and Kageki, Norri},
  journal={IEEE Robotics \& automation magazine},
  volume={19},
  number={2},
  pages={98--100},
  year={2012},
  publisher={IEEE}
}

@article{pfallerSocialPresenceModerator2021,
  title = {Social {{Presence}} as a {{Moderator}} of the {{Effect}} of {{Agent Behavior}} on {{Emotional Experience}} in {{Social Interactions}} in {{Virtual Reality}}},
  author = {Pfaller, Michael and Kroczek, Leon O. H. and Lange, Bastian and Fülöp, Raymund and Müller, Mathias and Mühlberger, Andreas},
  date = {2021-12-08},
  journaltitle = {Frontiers in Virtual Reality},
  shortjournal = {Front. Virtual Real.},
  volume = {2},
  pages = {741138},
  issn = {2673-4192},
  doi = {10.3389/frvir.2021.741138},
  url = {https://www.frontiersin.org/articles/10.3389/frvir.2021.741138/full},
  urldate = {2022-11-14},
  langid = {english}
}

@article{sprottAvatarsPhygitalSocial2025,
  title = {Avatars' {{Phygital Social Presence}} in the {{Metaverse}}: {{An Engaged Theory Perspective}}},
  shorttitle = {Avatars' Phygital Social Presence in the Metaverse: An Engaged Theory Perspective},
  author = {Sprott, David E. and Hollebeek, Linda D. and Sigurdsson, Valdimar and Clark, Moira K. and Urbonavicius, Sigitas},
  date = {2025-06},
  journaltitle = {Psychology \& Marketing},
  shortjournal = {Psychology and Marketing},
  volume = {42},
  number = {6},
  pages = {1528--1540},
  issn = {0742-6046, 1520-6793},
  doi = {10.1002/mar.22191},
  url = {https://onlinelibrary.wiley.com/doi/10.1002/mar.22191},
  urldate = {2025-09-19},
  langid = {english}
}

@article{bioccaMoreRobustTheory2003,
  title = {Toward a {{More Robust Theory}} and {{Measure}} of {{Social Presence}}: {{Review}} and {{Suggested Criteria}}},
  shorttitle = {Toward a {{More Robust Theory}} and {{Measure}} of {{Social Presence}}},
  author = {Biocca, Frank and Harms, Chad and Burgoon, Judee K.},
  date = {2003-10},
  journaltitle = {Presence: Teleoperators and Virtual Environments},
  shortjournal = {Presence: Teleoperators \& Virtual Environments},
  volume = {12},
  number = {5},
  pages = {456--480},
  issn = {1054-7460},
  doi = {10.1162/105474603322761270},
  url = {https://direct.mit.edu/pvar/article/12/5/456-480/58921},
  urldate = {2025-09-19},
  langid = {english}
}

@article{ohSystematicReviewSocial2018,
  title = {A {{Systematic Review}} of {{Social Presence}}: {{Definition}}, {{Antecedents}}, and {{Implications}}},
  shorttitle = {A {{Systematic Review}} of {{Social Presence}}},
  author = {Oh, Catherine S. and Bailenson, J. N. and Welch, Gregory F.},
  date = {2018-10-15},
  journaltitle = {Frontiers in Robotics and AI},
  shortjournal = {Front. Robot. AI},
  volume = {5},
  pages = {114},
  issn = {2296-9144},
  doi = {10.3389/frobt.2018.00114},
  url = {https://www.frontiersin.org/article/10.3389/frobt.2018.00114/full},
  urldate = {2022-12-14},
  langid = {english}
}

@article{tewellReviewOlfactoryDisplay2024,
  title = {A {{Review}} of {{Olfactory Display Designs}} for {{Virtual Reality Environments}}},
  author = {Tewell, Jordan and Ranasinghe, Nimesha},
  date = {2024-11-30},
  journaltitle = {ACM Computing Surveys},
  shortjournal = {ACM Comput. Surv.},
  volume = {56},
  number = {11},
  pages = {1--35},
  issn = {0360-0300, 1557-7341},
  doi = {10.1145/3665243},
  url = {https://dl.acm.org/doi/10.1145/3665243},
  urldate = {2025-09-19},
  langid = {english}
}

@article{oreganSensorimotorAccountVision2001,
  title = {A Sensorimotor Account of Vision and Visual Consciousness},
  author = {O'Regan, J. Kevin and Noë, Alva},
  date = {2001-10},
  journaltitle = {Behavioral and Brain Sciences},
  shortjournal = {Behav Brain Sci},
  volume = {24},
  number = {5},
  pages = {939--973},
  issn = {0140-525X, 1469-1825},
  doi = {10.1017/S0140525X01000115},
  url = {https://www.cambridge.org/core/product/identifier/S0140525X01000115/type/journal_article},
  urldate = {2025-09-19},
  langid = {english}
}

@inproceedings{fermoselleLetsGetTouch2020,
  title = {Let’s {{Get}} in {{Touch}}! {{Adding Haptics}} to {{Social VR}}},
  booktitle = {{{ACM International Conference}} on {{Interactive Media Experiences}}},
  author = {Fermoselle, Leonor and Gunkel, Simon and Ter Haar, Frank Ter and Dijkstra-Soudarissanane, Sylvie and Toet, Alexander and Niamut, Omar and Van Der Stap, Nanda Van},
  date = {2020-06-17},
  pages = {174--179},
  publisher = {ACM},
  location = {Cornella, Barcelona Spain},
  doi = {10.1145/3391614.3399396},
  url = {https://dl.acm.org/doi/10.1145/3391614.3399396},
  urldate = {2025-09-19},
  eventtitle = {{{IMX}} '20: {{ACM International Conference}} on {{Interactive Media Experiences}}},
  isbn = {978-1-4503-7976-2},
  langid = {english}
}

@article{slaterPlaceIllusionPlausibility2009,
  title = {Place Illusion and Plausibility Can Lead to Realistic Behaviour in Immersive Virtual Environments},
  author = {Slater, Mel},
  year = {2009},
  month = dec,
  journal = {Philosophical Transactions of the Royal Society B: Biological Sciences},
  volume = {364},
  number = {1535},
  pages = {3549--3557},
  issn = {0962-8436, 1471-2970},
  doi = {10.1098/rstb.2009.0138},
  urldate = {2025-09-19},
  langid = {english}
}

@article{cortesDelayThresholdSocial2024,
  title = {Delay {{Threshold}} for {{Social Interaction}} in {{Volumetric eXtended Reality Communication}}},
  author = {Cort{\'e}s, Carlos and Viola, Irene and Guti{\'e}rrez, Jes{\'u}s and Jansen, Jack and Subramanyam, Shishir and Alexiou, Evangelos and P{\'e}rez, Pablo and Garc{\'i}a, Narciso and C{\'e}sar, Pablo},
  year = {2024},
  month = jul,
  journal = {ACM Transactions on Multimedia Computing, Communications, and Applications},
  volume = {20},
  number = {7},
  pages = {1--22},
  issn = {1551-6857, 1551-6865},
  doi = {10.1145/3651164},
  urldate = {2025-09-19},
  langid = {english}
}

@article{slaterSeparateRealityUpdate2022,
  title = {A {{Separate Reality}}: {{An Update}} on {{Place Illusion}} and {{Plausibility}} in {{Virtual Reality}}},
  shorttitle = {A {{Separate Reality}}},
  author = {Slater, Mel and Banakou, Domna and Beacco, Alejandro and Gallego, Jaime and {Macia-Varela}, Francisco and Oliva, Ramon},
  year = {2022},
  month = jun,
  journal = {Frontiers in Virtual Reality},
  volume = {3},
  pages = {914392},
  issn = {2673-4192},
  doi = {10.3389/frvir.2022.914392},
  urldate = {2025-09-19}
}

@article{sonSocialVirtualReality2025,
  title = {Social Virtual Reality Elicits Natural Interaction Behavior with Self-Similar and Generic Avatars},
  author = {Son, Gayoung and Rubo, Marius},
  year = {2025},
  month = may,
  journal = {International Journal of Human-Computer Studies},
  volume = {199},
  pages = {103488},
  issn = {10715819},
  doi = {10.1016/j.ijhcs.2025.103488},
  urldate = {2025-08-11},
  langid = {english}
}

@article{aguertParaverbalExpressionVerbal2022,
  title = {Paraverbal {{Expression}} of {{Verbal Irony}}: {{Vocal Cues Matter}} and {{Facial Cues Even More}}},
  shorttitle = {Paraverbal {{Expression}} of {{Verbal Irony}}},
  author = {Aguert, Marc},
  date = {2022-03},
  journaltitle = {Journal of Nonverbal Behavior},
  shortjournal = {J Nonverbal Behav},
  volume = {46},
  number = {1},
  pages = {45--70},
  issn = {0191-5886, 1573-3653},
  doi = {10.1007/s10919-021-00385-z},
  url = {https://link.springer.com/10.1007/s10919-021-00385-z},
  urldate = {2025-08-11},
  langid = {english}
}

@article{seurenWhoseTurnIt2021,
  title = {Whose Turn Is It Anyway? {{Latency}} and the Organization of Turn-Taking in Video-Mediated Interaction},
  shorttitle = {Whose Turn Is It Anyway?},
  author = {Seuren, Lucas M. and Wherton, Joseph and Greenhalgh, Trisha and Shaw, Sara E.},
  date = {2021-01},
  journaltitle = {Journal of Pragmatics},
  shortjournal = {Journal of Pragmatics},
  volume = {172},
  pages = {63--78},
  issn = {03782166},
  doi = {10.1016/j.pragma.2020.11.005},
  url = {https://linkinghub.elsevier.com/retrieve/pii/S0378216620302782},
  urldate = {2025-08-11},
  langid = {english}
}

@article{kendrickTurntakingHumanFacetoface2023,
  title = {Turn-Taking in Human Face-to-Face Interaction Is Multimodal: Gaze Direction and Manual Gestures Aid the Coordination of Turn Transitions},
  shorttitle = {Turn-Taking in Human Face-to-Face Interaction Is Multimodal},
  author = {Kendrick, Kobin H. and Holler, Judith and Levinson, Stephen C.},
  date = {2023-04-24},
  journaltitle = {Philosophical Transactions of the Royal Society B: Biological Sciences},
  shortjournal = {Phil. Trans. R. Soc. B},
  volume = {378},
  number = {1875},
  pages = {20210473},
  issn = {0962-8436, 1471-2970},
  doi = {10.1098/rstb.2021.0473},
  url = {https://royalsocietypublishing.org/doi/10.1098/rstb.2021.0473},
  urldate = {2025-08-11},
  langid = {english}
}

@book{sebeok2001signs,
  title={Signs: An introduction to semiotics},
  author={Sebeok, Thomas},
  year={2001},
  publisher={University of Toronto Press}
}

@article{ovsyannikova2025,
  title = {Third-Party Evaluators Perceive {{AI}} as More Compassionate than Expert Humans},
  author = {Ovsyannikova, Dariya and De Mello, Victoria Oldemburgo and Inzlicht, Michael},
  date = {2025-01-10},
  journaltitle = {Communications Psychology},
  shortjournal = {Commun Psychol},
  volume = {3},
  number = {1},
  pages = {4},
  issn = {2731-9121},
  doi = {10.1038/s44271-024-00182-6},
  url = {https://www.nature.com/articles/s44271-024-00182-6},
  urldate = {2025-02-27},
  langid = {english}
}

@inproceedings{tsigeman2024psychological,
  title={Psychological Aspects of Face-To-Face Versus Computer-Mediated Interpersonal Communication: An Integrative Review},
  author={Tsigeman, Elina and Mararitsa, Larisa and Gundelah, Olga and Lopatina, Olga and Koltsova, Olessia},
  booktitle={International Conference on Human-Computer Interaction},
  pages={29--48},
  year={2024},
  organization={Springer}
}

@article{fuss2019computer,
  title={Computer-mediated communication and social support among community-dwelling older adults: A systematic review of cross-sectional data},
  author={Fuss, Belinda G and Dorstyn, Diana and Ward, Lynn},
  journal={Australasian journal on ageing},
  volume={38},
  number={4},
  pages={e103--e113},
  year={2019},
  publisher={Wiley Online Library}
}

@article{yao2020computer,
  title={“{What} is computer-mediated communication?”—An introduction to the special issue},
  author={Yao, Mike Z and Ling, Rich},
  journal={Journal of Computer-Mediated Communication},
  volume={25},
  number={1},
  pages={4--8},
  year={2020},
  publisher={Oxford University Press},
  doi={10.1093/jcmc/zmz027}
}

@article{luong2011better,
  title={Better with age: Social relationships across adulthood},
  author={Luong, Gloria and Charles, Susan T and Fingerman, Karen L},
  journal={Journal of Social and Personal Relationships},
  volume={28},
  number={1},
  pages={9--23},
  year={2011},
  publisher={Sage Publications Sage UK: London, England},
  doi={10.1177/0265407510391362}
}

@article{hulur2020rethinking,
  title={Rethinking social relationships in old age: Digitalization and the social lives of older adults.},
  author={H{\"u}l{\"u}r, Gizem and Macdonald, Birthe},
  journal={American Psychologist},
  volume={75},
  number={4},
  pages={554},
  year={2020},
  publisher={American Psychological Association},
  doi={10.1037/amp0000604}
}

@article{vaportzis2017older,
  title={Older adults perceptions of technology and barriers to interacting with tablet computers: a focus group study},
  author={Vaportzis, Eleftheria and Giatsi Clausen, Maria and Gow, Alan J},
  journal={Frontiers in psychology},
  volume={8},
  pages={1687},
  year={2017},
  publisher={Frontiers Media SA}
}

@ARTICLE{Gannot2017,
  author={Gannot, Sharon and Vincent, Emmanuel and Markovich-Golan, Shmulik and Ozerov, Alexey},
  journal={IEEE/ACM Transactions on Audio, Speech, and Language Processing}, 
  title={A Consolidated Perspective on Multimicrophone Speech Enhancement and Source Separation}, 
  year={2017},
  volume={25},
  number={4},
  pages={692-730},
  doi={10.1109/TASLP.2016.2647702}}

@article{Frith2012,
  title = {Mechanisms of {{Social Cognition}}},
  author = {Frith, C. D. and Frith, Uta},
  date = {2012},
  journaltitle = {Annual review of Psychology},
  volume = {63},
  eprint = {21838544},
  eprinttype = {pubmed},
  pages = {287--313},
  issn = {0066-4308},
  doi = {10.1146/annurev-psych-120710-100449},
  OPTisbn = {0066-4308\textbackslash r978-0-8243-0263-4}
}

@book{argyleBodilyCommunication2013,
  title = {Bodily {{Communication}}},
  author = {Argyle, Michael},
  date = {2013},
  edition = {2},
  publisher = {Routledge},
  location = {London},
  doi = {10.4324/9780203753835},
  url = {https://www.taylorfrancis.com/books/9781134964253},
  urldate = {2025-04-04},
  isbn = {978-1-134-96425-3},
  langid = {english}
}

@article{Holle2007,
  title = {The Role of Iconic Gestures in Speech Disambiguation: {{ERP}} Evidence},
  author = {Holle, Henning and Gunter, Thomas C},
  date = {2007},
  journaltitle = {Journal of Cognitive Neuroscience},
  volume = {19},
  number = {7},
  pages = {1175--1192},
  doi={10.1162/jocn.2007.19.7.1175}
}

@article{wohltjenEyeContactMarks2021,
  title = {Eye Contact Marks the Rise and Fall of Shared Attention in Conversation},
  author = {Wohltjen, Sophie and Wheatley, Thalia},
  date = {2021-09-14},
  journaltitle = {Proceedings of the National Academy of Sciences},
  shortjournal = {Proc. Natl. Acad. Sci. U.S.A.},
  volume = {118},
  number = {37},
  pages = {e2106645118},
  issn = {0027-8424, 1091-6490},
  doi = {10.1073/pnas.2106645118},
  url = {https://pnas.org/doi/full/10.1073/pnas.2106645118},
  urldate = {2025-04-04},
  langid = {english}
}

@incollection{matsumotoBodyPosturesGait2016,
  title = {The Body: {{Postures}}, Gait, Proxemics, and Haptics.},
  shorttitle = {The Body},
  booktitle = {{{APA}} Handbook of Nonverbal Communication.},
  author = {Matsumoto, David and Hwang, Hyisung C. and Frank, Mark G.},
  editor = {Matsumoto, David and Hwang, Hyisung C. and Frank, Mark G.},
  date = {2016},
  pages = {387--400},
  publisher = {American Psychological Association},
  location = {Washington},
  doi = {10.1037/14669-015},
  url = {https://content.apa.org/books/14669-015},
  urldate = {2025-04-04},
  isbn = {978-1-4338-1969-8},
  langid = {english}
}

@article{bohannonEyeContactVideomediated2013,
  title = {Eye Contact and Video-Mediated Communication: {{A}} Review},
  shorttitle = {Eye Contact and Video-Mediated Communication},
  author = {Bohannon, Leanne S. and Herbert, Andrew M. and Pelz, Jeff B. and Rantanen, Esa M.},
  date = {2013-04},
  journaltitle = {Displays},
  shortjournal = {Displays},
  volume = {34},
  number = {2},
  pages = {177--185},
  issn = {01419382},
  doi = {10.1016/j.displa.2012.10.009},
  url = {https://linkinghub.elsevier.com/retrieve/pii/S0141938212001084},
  urldate = {2025-04-04},
  langid = {english}
}

@article{langeReadingEmotionsReading2022,
  title = {Reading Emotions, Reading People: {{Emotion}} Perception and Inferences Drawn from Perceived Emotions},
  shorttitle = {Reading Emotions, Reading People},
  author = {Lange, Jens and Heerdink, Marc W. and Van Kleef, Gerben A.},
  date = {2022-02},
  journaltitle = {Current Opinion in Psychology},
  shortjournal = {Current Opinion in Psychology},
  volume = {43},
  pages = {85--90},
  issn = {2352250X},
  doi = {10.1016/j.copsyc.2021.06.008},
  url = {https://linkinghub.elsevier.com/retrieve/pii/S2352250X21000774},
  urldate = {2024-03-28},
  langid = {english}
}

@article{vankleefHowEmotionsRegulate2009,
  title = {How {{Emotions Regulate Social Life}}: {{The Emotions}} as {{Social Information}} ({{EASI}}) {{Model}}},
  shorttitle = {How {{Emotions Regulate Social Life}}},
  author = {Van Kleef, Gerben A.},
  date = {2009-06},
  journaltitle = {Current Directions in Psychological Science},
  shortjournal = {Curr Dir Psychol Sci},
  volume = {18},
  number = {3},
  pages = {184--188},
  issn = {0963-7214, 1467-8721},
  doi = {10.1111/j.1467-8721.2009.01633.x},
  url = {http://journals.sagepub.com/doi/10.1111/j.1467-8721.2009.01633.x},
  urldate = {2024-03-28},
  langid = {english}
}

@article{frithTheoryMind2005,
  title = {Theory of Mind},
  author = {Frith, C. D. and Frith, Uta},
  date = {2005-09},
  journaltitle = {Current Biology},
  shortjournal = {Current Biology},
  volume = {15},
  number = {17},
  pages = {R644-R645},
  issn = {09609822},
  doi = {10.1016/j.cub.2005.08.041},
  url = {https://linkinghub.elsevier.com/retrieve/pii/S0960982205009607},
  urldate = {2025-04-04},
  langid = {english}
}

@article{danielc.richardsonArtConversationCoordination2007,
  title = {The {{Art}} of {{Conversation Is Coordination Common Ground}} and the {{Coupling}} of {{Eye Movements During Dialogue}}},
  author = {Richardson, Daniel C. and 
      Dale, Rick and Kirkham, Natasha Z.},
  date = {2007-05-01},
  journaltitle = {Psychological Science},
  volume = {18},
  number = {5},
  eprint = {17576280},
  eprinttype = {pubmed},
  pages = {407--413},
  doi = {10.1111/j.1467-9280.2007.01914.x}
}

@article{hadleyReviewTheoriesMethods2022,
  title = {A Review of Theories and Methods in the Science of Face-to-Face Social Interaction},
  author = {Hadley, Lauren V. and Naylor, Graham and Hamilton, Antonia F. De C.},
  date = {2022-01-12},
  journaltitle = {Nature Reviews Psychology},
  shortjournal = {Nat Rev Psychol},
  volume = {1},
  number = {1},
  pages = {42--54},
  issn = {2731-0574},
  doi = {10.1038/s44159-021-00008-w},
  url = {https://www.nature.com/articles/s44159-021-00008-w},
  urldate = {2023-06-20},
  langid = {english}
}

@inproceedings{ehretWhosNextIntegrating2023,
  title = {Who's next?: {{Integrating Non-Verbal Turn-Taking Cues}} for {{Embodied Conversational Agents}}},
  shorttitle = {Who's Next?},
  booktitle = {Proceedings of the 23rd {{ACM International Conference}} on {{Intelligent Virtual Agents}}},
  author = {Ehret, Jonathan and Bönsch, Andrea and Nossol, Patrick and Ermert, Cosima A. and Mohanathasan, Chinthusa and Schlittmeier, Sabine J. and Fels, Janina and Kuhlen, Torsten W.},
  date = {2023-09-19},
  pages = {1--8},
  publisher = {ACM},
  location = {Würzburg Germany},
  doi = {10.1145/3570945.3607312},
  url = {https://dl.acm.org/doi/10.1145/3570945.3607312},
  urldate = {2025-04-04},
  eventtitle = {{{IVA}} '23: {{ACM International Conference}} on {{Intelligent Virtual Agents}}},
  isbn = {978-1-4503-9994-4},
  langid = {english}
}

@article{nassMachinesMindlessnessSocial2000,
  title = {Machines and {{Mindlessness}}: {{Social Responses}} to {{Computers}}},
  shorttitle = {Machines and {{Mindlessness}}},
  author = {Nass, Clifford and Moon, Youngme},
  date = {2000-01},
  journaltitle = {Journal of Social Issues},
  shortjournal = {Journal of Social Issues},
  volume = {56},
  number = {1},
  pages = {81--103},
  issn = {0022-4537, 1540-4560},
  doi = {10.1111/0022-4537.00153},
  urldate = {2024-08-09},
  langid = {english}
}

@article{kroczekInfluencePersonaConversational2025,
  title = {The Influence of Persona and Conversational Task on Social Interactions with a {{LLM-controlled}} Embodied Conversational Agent},
  author = {Kroczek, Leon O. H. and May, Alexander and Hettenkofer, Selina and Ruider, Andreas and Ludwig, Bernd and Mühlberger, Andreas},
  date = {2025-07},
  journaltitle = {Computers in Human Behavior},
  shortjournal = {Computers in Human Behavior},
  pages = {108759},
  issn = {07475632},
  doi = {10.1016/j.chb.2025.108759},
  url = {https://linkinghub.elsevier.com/retrieve/pii/S0747563225002067},
  urldate = {2025-08-01},
  langid = {english}
}

@incollection{watzlawick2017tentative,
  author    = {Watzlawick, Paul and Beavin, Janet and Jackson, Don D.},
  year      = {2017},
  title     = {Some tentative axioms of communication},
  booktitle = {Communication Theory},
  pages     = {74--80},
  publisher = {Routledge},
  address   = {New York}
}

@inproceedings{pichora2015hearing,
  title={Hearing, cognition, and healthy aging: Social and public health implications of the links between age-related declines in hearing and cognition},
  author={Pichora-Fuller, M Kathleen and Mick, Paul and Reed, Marilyn},
  booktitle={Seminars in hearing},
  volume={36},
  number={03},
  pages={122--139},
  year={2015},
  organization={Thieme Medical Publishers}
}

@article{irfan2024recommendations,
  title={Recommendations for designing conversational companion robots with older adults through foundation models},
  author={Irfan, Bahar and Kuoppam{\"a}ki, Sanna and Skantze, Gabriel},
  journal={Frontiers in Robotics and AI},
  volume={11},
  pages={1363713},
  year={2024},
  publisher={Frontiers Media SA},
  doi={/10.3389/frobt.2024.1363713}
}

@article{yang2024talk2care,
  title={Talk2care: An {LLM}-based voice assistant for communication between healthcare providers and older adults},
  author={Yang, Ziqi and Xu, Xuhai and Yao, Bingsheng and Rogers, Ethan and Zhang, Shao and Intille, Stephen and Shara, Nawar and Gao, Guodong Gordon and Wang, Dakuo},
  journal={Proceedings of the ACM on Interactive, Mobile, Wearable and Ubiquitous Technologies},
  volume={8},
  number={2},
  pages={1--35},
  year={2024},
  publisher={ACM New York, NY, USA}
}

@article{chen2024scoping,
  title={A SCOPING REVIEW OF THE ARTIFICIAL INTELLIGENCE--BASED CONVERSATIONAL AGENTS ON MENTAL HEALTH CARE FOR OLDER ADULTS},
  author={Chen, Xiayu and Wen, Wan},
  journal={Innovation in Aging},
  volume={8},
  number={Supplement\_1},
  pages={257--257},
  year={2024},
  publisher={Oxford University Press US},
  doi={10.1093/geroni/igae098.0831}
}

@article{Livingston2024,
  title={Dementia prevention, intervention, and care: 2024 report of the {Lancet} standing Commission},
  author={Gill Livingston and Jonathan Huntley and Kathy Y Liu and Sergi G Costafreda and Geir Selbæk and Suvarna Alladi and  David Ames and  Sube Banerjee and  Alistair Burns and  Carol Brayne and  Nick C Fox and  Cleusa P Ferri and  Laura N Gitlin and  Robert Howard and  Helen C Kales and  Mika Kivimäki and  Eric B Larson and  Noeline Nakasujja and  Kenneth Rockwood and  Quincy Samus and  Kokoro Shirai and  Archana Singh-Manoux and  Lon S Schneider and  Sebastian Walsh and  Yao Yao and  Andrew Sommerlad and Naaheed Mukadam},
  journal={The Lancet},
  volume={404},
  issue={10452},
  OPTnumber={},
  pages={572–628},
  year={2024},
  OPTpublisher={}
}

@article{Cherry1953,
  title={Some experiments on the recognition of speech, with one and with two ears},
  author={Cherry, E C},
  journal={Journal of the Acoustical Society of America},
  volume={25},
  OPTnumber={},
  pages={975–979},
  year={1953},
  OPTpublisher={}
}

@book{bailenson2018experience,
  title={Experience on demand: What virtual reality is, how it works, and what it can do},
  author={Bailenson, J. N.},
  year={2018},
  publisher={WW Norton \& Company}
}

@book{dignum2019responsible,
  author    = {Virginia Dignum},
  title     = {Responsible Artificial Intelligence: How to Develop and Use {AI} in a Responsible Way},
  publisher = {Springer Nature},
  year      = {2019}
}

@inproceedings{gratch2007can,
  author    = {Jonathan Gratch and Ning Wang and Anna Okhmatovskaia and Frédéric Lamothe and Mathieu Morales and Richard J. van der Werf and Louis-Philippe Morency},
  title     = {Can Virtual Humans Be More Engaging Than Real Ones?},
  booktitle = {Intelligent Virtual Agents: 7th International Conference, IVA 2007, Paris, France, September 17-19, 2007. Proceedings},
  pages     = {254--267},
  year      = {2007},
  publisher = {Springer Berlin Heidelberg}
}

@inproceedings{hovy2016social,
  author    = {Dirk Hovy and Shannon L. Spruit},
  title     = {The Social Impact of Natural Language Processing},
  booktitle = {Proceedings of the 54th Annual Meeting of the Association for Computational Linguistics (Volume 2: Short Papers)},
  pages     = {591--598},
  year      = {2016}
}

@article{koenecke2020racial,
  author    = {Allison Koenecke and Andrew Nam and Emily Lake and Joe Nudell and Mikhail Quartey and Zaid Mengesha and ... and Sharad Goel},
  title     = {Racial Disparities in Automated Speech Recognition},
  journal   = {Proceedings of the National Academy of Sciences},
  volume    = {117},
  number    = {14},
  pages     = {7684--7689},
  year      = {2020}
}

@article{lundberg2017unified,
  author    = {Scott M. Lundberg and Su-In Lee},
  title     = {A Unified Approach to Interpreting Model Predictions},
  journal   = {Advances in Neural Information Processing Systems},
  volume    = {30},
  year      = {2017}
}

@article{nissenbaum2011contextual,
  author    = {Helen Nissenbaum},
  title     = {A Contextual Approach to Privacy Online},
  journal   = {Daedalus},
  volume    = {140},
  number    = {4},
  pages     = {32--48},
  year      = {2011},
  doi = {10.1162/DAED_a_00113}
}

@article{nosek2015promoting,
  author    = {Brian A. Nosek and George Alter and George C. Banks and Denny Borsboom and Sara D. Bowman and Steven J. Breckler and ... and Tal Yarkoni},
  title     = {Promoting an Open Research Culture},
  journal   = {Science},
  volume    = {348},
  number    = {6242},
  pages     = {1422--1425},
  year      = {2015},
  doi    = {10.1126/science.aab2374}
}

@book{voigt2017eu,
  author    = {Paul Voigt and Axel von dem Bussche},
  title     = {The {EU} General Data Protection Regulation ({GDPR}): A Practical Guide},
  publisher = {Springer},
  year      = {2017}
}

@article{getzmann2017visually,
  title={Visually guided auditory attention in a dynamic “cocktail-party” speech perception task: {ERP} evidence for age-related differences},
  author={Getzmann, Stephan and Wascher, Edmund},
  journal={Hearing Research},
  volume={344},
  pages={98--108},
  year={2017},
  publisher={Elsevier},
  doi={10.1016/j.heares.2016.11.001}
}

@article{begau2022role,
  title={The role of informational content of visual speech in an audiovisual cocktail party: Evidence from cortical oscillations in young and old participants},
  author={Begau, Alexandra and Klatt, Laura-Isabelle and Schneider, Daniel and Wascher, Edmund and Getzmann, Stephan},
  journal={European Journal of Neuroscience},
  volume={56},
  number={8},
  pages={5215--5234},
  year={2022},
  publisher={Wiley Online Library},
  doi = {10.1111/ejn.15811}
}

@article{sohn1999statistical,
  title={A statistical model-based voice activity detection},
  author={Sohn, J. and Kim, N. S. and Sung, W.},
  journal={IEEE Signal Processing Letters},
  volume={6},
  number={1},
  pages={1--3},
  year={1999},
  publisher={IEEE}
}

@article{heldner2010pauses,
  title={Pauses, gaps and overlaps in conversations},
  author={Heldner, M. and Edlund, J.},
  journal={Journal of Phonetics},
  volume={38},
  number={4},
  pages={555--568},
  year={2010},
  publisher={Elsevier},
  doi={10.1016/j.wocn.2010.08.002}
}

@book{levinson1983pragmatics,
  title={Pragmatics},
  author={Levinson, S. C.},
  year={1983},
  publisher={Cambridge University Press}
}

@incollection{couperkuhlen2001interactional,
  title={Interactional prosody: High-onsets in theory and practice},
  author={Couper-Kuhlen, E.},
  booktitle={Studies in interactional linguistics},
  editor={Selting, D. and Couper-Kuhlen, M.},
  pages={99--128},
  year={2001},
  publisher={John Benjamins}
}

@article{kendon1967some,
  title={Some functions of gaze-direction in social interaction},
  author={Kendon, A.},
  journal={Acta Psychologica},
  volume={26},
  pages={22--63},
  year={1967},
  publisher={Elsevier}
}

@article{picton2000guidelines,
  title={Guidelines for using human event-related potentials to study cognition: Recording standards and publication criteria},
  author={Picton, T. W. and Bentin, S. and Berg, P. and Donchin, E. and Hillyard, S. A. and Johnson, R. and Miller, G. A. and Ritter, W. and Ruchkin, D. S. and Rugg, M. D. and Taylor, M. J.},
  journal={Psychophysiology},
  volume={37},
  number={2},
  pages={127--152},
  year={2000},
  publisher={Wiley Online Library},
  doi={10.1111/1469-8986.3720127}
}

@article{babiloni2014social,
  title={Social neuroscience and hyperscanning techniques: Past, present and future},
  author={Babiloni, F. and Astolfi, L.},
  journal={Neuroscience \& Biobehavioral Reviews},
  volume={44},
  pages={76--93},
  year={2014},
  publisher={Elsevier},
  doi={10.1016/j.neubiorev.2012.07.006}
}

@article{luo2007phase,
  title={Phase patterns of neuronal responses reliably discriminate speech in human auditory cortex},
  author={Luo, H. and Poeppel, D.},
  journal={Neuron},
  volume={54},
  number={6},
  pages={1001--1010},
  year={2007},
  publisher={Elsevier}
}

@article{foxe2011role,
  title={The role of alpha-band brain oscillations as a sensory suppression mechanism during selective attention},
  author={Foxe, J. J. and Snyder, A. C.},
  journal={Frontiers in Psychology},
  volume={2},
  pages={154},
  year={2011},
  publisher={Frontiers},
  doi={10.3389/fpsyg.2011.00154}
}

@article{arnal2012cortical,
  title={Cortical oscillations and speech processing},
  author={Arnal, L. H. and Giraud, A. L.},
  journal={Trends in cognitive sciences},
  volume={16},
  number={7},
  pages={390--398},
  year={2012},
  publisher={Elsevier},
  doi = {10.1038/nn.3063}
}

@article{hasson2012brain,
  title={Brain-to-brain coupling: a mechanism for creating and sharing a social world},
  author={Hasson, U. and Ghazanfar, A. A. and Galantucci, B. and Garrod, S. and Keysers, C.},
  journal={Trends in Cognitive Sciences},
  volume={16},
  number={2},
  pages={114--121},
  year={2012},
  publisher={Elsevier},
  doi={10.1016/j.tics.2011.12.007}
}

@article{crosse2016multivariate,
  title={The multivariate temporal response function (mTRF) toolbox: a MATLAB toolbox for relating neural signals to continuous stimuli},
  author={Crosse, M. J. and Di Liberto, G. M. and Bednar, A. and Lalor, E. C.},
  journal={Frontiers in Human Neuroscience},
  volume={10},
  pages={604},
  year={2016},
  publisher={Frontiers},
  doi={10.3389/fnhum.2016.00604}
}

@article{fridlund1986guidelines,
  title={Guidelines for human electromyographic research},
  author={Fridlund, A. J. and Cacioppo, J. T.},
  journal={Psychophysiology},
  volume={23},
  number={5},
  pages={567--589},
  year={1986},
  publisher={Wiley Online Library}
}

@article{hickok2012computational,
  title={Computational neuroanatomy of speech production},
  author={Hickok, G.},
  journal={Nature Reviews Neuroscience},
  volume={13},
  number={2},
  pages={135--145},
  year={2012},
  publisher={Nature Publishing Group},
  doi={10.1038/nrn3158}
}

@article{hess2013emotional,
  title={Emotional mimicry as social regulation},
  author={Hess, U. and Fischer, A.},
  journal={Personality and Social Psychology Review},
  volume={17},
  number={2},
  pages={142--157},
  year={2013},
  publisher={Sage Publications Sage CA: Los Angeles, CA},
  doi={10.1177/1088868312472607}
}

@article{cacioppo1986electromyographic,
  title={Electromyographic activity over facial muscle regions can differentiate the valence and intensity of affective reactions},
  author={Cacioppo, J. T. and Petty, R. E. and Losch, M. E. and Kim, H. S.},
  journal={Journal of Personality and Social Psychology},
  volume={50},
  number={2},
  pages={260--268},
  year={1986},
  publisher={American Psychological Association}
}

@article{palumbo2017interpersonal,
  title={Interpersonal autonomic physiology: A systematic review of the literature},
  author={Palumbo, R. V. and Marraccini, M. E. and Weyandt, L. L. and Wilder-Smith, O. and McGee, H. A. and Liu, S. and Goodwin, M. S.},
  journal={Personality and Social Psychology Review},
  volume={21},
  number={2},
  pages={99--141},
  year={2017},
  publisher={Sage Publications Sage CA: Los Angeles, CA},
  doi={10.1177/1088868316628405}
}

@incollection{giles1991accommodation,
  title={Accommodation theory: Communication, context, and consequence},
  author={Giles, H. and Coupland, N. and Coupland, J.},
  booktitle={Contexts of accommodation: Developments in applied sociolinguistics},
  editor={Giles, H. and Coupland, J. and Coupland, N.},
  pages={1--68},
  year={1991},
  publisher={Cambridge University Press}
}

@article{tickle1990nature,
  title={The nature of rapport and its nonverbal correlates},
  author={Tickle-Degnen, L. and Rosenthal, R.},
  journal={Psychological Inquiry},
  volume={1},
  number={4},
  pages={285--293},
  year={1990},
  publisher={Taylor \& Francis}
}

@article{chartrand1999chameleon,
  title={The chameleon effect: The perception--behavior link and social interaction},
  author={Chartrand, T. L. and Bargh, J. A.},
  journal={Journal of Personality and Social Psychology},
  volume={76},
  number={6},
  pages={893--910},
  year={1999},
  publisher={American Psychological Association}
}

@article{pardo2006phonetic,
  title={On phonetic convergence during conversational interaction},
  author={Pardo, J. S.},
  journal={The Journal of the Acoustical Society of America},
  volume={119},
  number={4},
  pages={2382--2393},
  year={2006},
  publisher={ASA}
}

@article{boersma2001praat,
  title={Praat, a system for doing phonetics by computer},
  author={Boersma, P. and Weenink, D.},
  journal={Glot international},
  volume={5},
  number={9/10},
  pages={341--345},
  year={2001}
}

@inproceedings{eyben2010opensmile,
  title={Opensmile: The {mMnich} versatile and fast open-source audio feature extractor},
  author={Eyben, F. and W{\"o}llmer, M. and Schuller, B.},
  booktitle={Proceedings of the 18th ACM international conference on Multimedia},
  pages={1459--1462},
  year={2010},
  organization={ACM}
}

@article{rochetcapellan2014take,
  title={Take a breath and take the turn: How breathing meets turns in spontaneous dialogue},
  author={Rochet-Capellan, A. and Fuchs, S.},
  journal={Philosophical Transactions of the Royal Society B: Biological Sciences},
  volume={369},
  number={1658},
  pages={20130399},
  year={2014},
  publisher={The Royal Society},
  doi={10.1098/rstb.2013.0399}
}

@article{louwerse2012behavior,
  title={Behavior matching in multimodal communication is evidence for shared conceptual representations},
  author={Louwerse, M. M. and Dale, R. and Bard, E. G. and Jeuniaux, P.},
  journal={Cognitive science},
  volume={36},
  number={8},
  pages={1405--1426},
  year={2012},
  publisher={Wiley Online Library}
}

@article{walther1996computer,
  title={Computer-mediated communication: Impersonal, interpersonal, and hyperpersonal interaction},
  author={Walther, J. B.},
  journal={Communication research},
  volume={23},
  number={1},
  pages={3--43},
  year={1996},
  publisher={Sage Publications Sage CA: Thousand Oaks, CA}
}

@article{cui2012nirs,
  title={{NIRS}-based hyperscanning reveals increased interpersonal coherence in superior frontal cortex during cooperation},
  author={Cui, X. and Bryant, D. M. and Reiss, A. L.},
  journal={Neuroimage},
  volume={59},
  number={3},
  pages={2430--2437},
  year={2012},
  publisher={Elsevier}
}

@InProceedings{radford2023robust,
  title = 	 {Robust Speech Recognition via Large-Scale Weak Supervision},
  author =       {Radford, Alec and Kim, Jong Wook and Xu, Tao and Brockman, Greg and Mcleavey, Christine and Sutskever, Ilya},
  booktitle = 	 {Proceedings of the 40th International Conference on Machine Learning},
  pages = 	 {28492--28518},
  year = 	 {2023},
  editor = 	 {Krause, Andreas and Brunskill, Emma and Cho, Kyunghyun and Engelhardt, Barbara and Sabato, Sivan and Scarlett, Jonathan},
  volume = 	 {202},
  series = 	 {Proceedings of Machine Learning Research},
  month = 	 {23--29 Jul},
  publisher =    {PMLR},
  url = 	 {https://proceedings.mlr.press/v202/radford23a.html}
}

@inproceedings{Marsh2023,
  author    = {Marsh, J.E. and Gädtke, J.F. and Schlittmeier, S.J.},
  title     = {A Review of the Effect of Noise on Cognitive Performance 2021-2023},
  booktitle = {13th ICBEN Congress on Noise as a Public Health Problem},
  year      = {2023},
  month     = {6},
  address   = {Belgrade, Serbia},
  url       = {https://www.icben.org/2023/presenting190.pdf}
}

@article{Azari2024,
  title={The prevalence of voice disorders and the related factors in university professors: a systematic review and meta-analysis},
  author={Azari, Sanaz and Aghaz, Alireza and Maarefvand, Mohammad and Ghelichi, Leila and Pashazadeh, Fariba and Shavaki, Younes Amiri},
  journal={Journal of Voice},
  volume={38},
  number={5},
  pages={1103--1114},
  year={2024},
  publisher={Elsevier},
  doi = {10.1016/j.jvoice.2022.02.017}
}

@article{Schiller2023,
  title={The impact of a speaker’s voice quality on auditory perception and cognition: A behavioral and subjective approach},
  author={Schiller, Isabel S and Asp{\"o}ck, Lukas and Schlittmeier, Sabine J},
  journal={Frontiers in Psychology},
  volume={14},
  pages={1243249},
  year={2023},
  publisher={Frontiers Media SA},
  doi={10.3389/fpsyg.2023.1243249}
}

@article{Schiller2024,
  title={A lecturer’s voice quality and its effect on memory, listening effort, and perception in a {VR} environment},
  author={Schiller, Isabel S and Breuer, Carolin and Asp{\"o}ck, Lukas and Ehret, Jonathan and B{\"o}nsch, Andrea and Kuhlen, Torsten W and Fels, Janina and Schlittmeier, Sabine J},
  journal={Scientific Reports},
  volume={14},
  number={1},
  pages={12407},
  year={2024},
  publisher={Nature Publishing Group UK London},
  doi={10.1038/s41598-024-63097-6}
}

@Book{Vary24,
  author = {P. Vary and R. Martin},
  ALTeditor = 	 {},
  title = 	 {Digital Speech Transmission and Enhancement},
  publisher = 	 {John Wiley \& Sons / IEEE Press},
  year = 	 {2024},
  OPTkey = 	 {},
  OPTvolume = 	 {},
  OPTnumber = 	 {},
  OPTseries = 	 {},
  address = 	 {Chichester},
  edition = 	 {2nd}
}

@misc{meta-metaverse,
    title = {What is the metaverse},
    author = {Meta},
    year = {2025},
     urldate = {2025},
    OPTkey = {},
    url = {https://www.meta.com/de-de/metaverse/what-is-the-metaverse/?utm_source=about.facebook.com&utm_medium=redirect}
}

@misc{secondlife,
    author = {{Linden Lab}},
    year = {2025},
    urldate = {2025},
    OPTkey = {},
    url = {https://lindenlab.com}
}

@misc{microsoftmesh,
    author = {{Microsoft}},
    year = {2025},
    urldate = {2025},
    OPTkey = {},
    url = {https://www.microsoft.com/en-us/microsoft-teams/microsoft-mesh}
}

@misc{teamviewer,
    author = {Teamviewer},
year  = {2025},
    OPTnote = {},
url ={https://www.teamviewer.com/ams/insights/the-metaverse-is-for-real/},
urldate ={2025}
}

@book{Blauert1996,
   author = {Jens Blauert},
   doi = {10.7551/mitpress/6391.001.0001},
   isbn = {9780262268684},
   month = {10},
   publisher = {The MIT Press},
   title = {Spatial Hearing},
   year = {1996},
}

@book{Vorlaender08,
    author = {Michael Vorländer},
    title = {Auralization --
Fundamentals of Acoustics, Modelling, Simulation, Algorithms and Acoustic Virtual Reality},
    publisher = {Springer Berlin, Heidelberg},
    doi ={10.1007/978-3-540-48830-9},
    year = {2008}
}

@article{wendt2014a,
		title = {A Computationally-Efficient and Perceptually-Plausible Algorithm for Binaural Room Impulse Response Simulation},
		author = {Wendt, Torben and van de Par, Steven and Ewert, Stephan D.},
		journal = {Journal of the Audio Engineering Society},
		volume = {62},
		pages = {748--766},
		year = {2014},
		url = {https://www.aes.org/e-lib/browse.cfm?elib=17550}
		}

@inproceedings{Gergen2012,
   author = {Sebastian Gergen and Christian Borss and Nilesh Madhu and Rainer Martin},
   doi = {10.1109/ICSPCC.2012.6335733},
   isbn = {9781467321938},
   booktitle = {IEEE International Conference on Signal Processing, Communications and Computing, ICSPCC},
   pages = {154-157},
   title = {An optimized parametric model for the simulation of reverberant microphone signals},
   year = {2012},
   pages = {3}
}

@INPROCEEDINGS{Zohourian16,
  author={Zohourian, Mehdi and Martin, Rainer},
  booktitle={2016 IEEE International Conference on Acoustics, Speech and Signal Processing (ICASSP)}, 
  title={Binaural speaker localization and separation based on a joint {ITD/ILD} model and head movement tracking}, 
  year={2016},
  volume={},
  number={},
  pages={430-434},
  doi={10.1109/ICASSP.2016.7471711}}

@Book{Kates2008,
  editor       = {James M. Kates},
  publisher    = {Plural Pub},
  title        = {Digital hearing aids},
  year         = {2008},
  address      = {San Diego},
  isbn         = {1597568333},
  creationdate = {2025-02-11T15:48:31},
  pagetotal    = {449},
  ppn_gvk      = {834364506},
}

@Article{Doclo2015,
  author       = {Doclo, Simon and Kellermann, Walter and Makino, Shoji and Nordholm, Sven Erik},
  journal      = {IEEE Signal Processing Magazine},
  title        = {Multichannel Signal Enhancement Algorithms for Assisted Listening Devices: Exploiting spatial diversity using multiple microphones},
  year         = {2015},
  issn         = {1558-0792},
  month        = mar,
  number       = {2},
  pages        = {18--30},
  volume       = {32},
  creationdate = {2025-02-28T10:23:10},
  doi          = {10.1109/msp.2014.2366780},
  publisher    = {Institute of Electrical and Electronics Engineers (IEEE)},
  readstatus   = {skimmed},
}

@Article{Kortlang2017,
  author     = {Kortlang, Steffen and Chen, Zhangli and Gerkmann, Timo and Kollmeier, Birger and Hohmann, Volker and Ewert, Stephan D.},
  journal    = {International Journal of Audiology},
  title      = {Evaluation of combined dynamic compression and single channel noise reduction for hearing aid applications},
  year       = {2017},
  issn       = {1708-8186},
  month      = mar,
  number     = {sup3},
  pages      = {S43--S54},
  volume     = {57},
  doi        = {10.1080/14992027.2017.1300695},
  publisher  = {Informa UK Limited},
  readstatus = {read},
}

@Article{May2018,
  author       = {May, Tobias and Kowalewski, Borys and Dau, Torsten},
  journal      = {Trends in Hearing},
  title        = {Signal-to-Noise-Ratio-Aware Dynamic Range Compression in Hearing Aids},
  year         = {2018},
  issn         = {2331-2165},
  month        = jan,
  volume       = {22},
  creationdate = {2025-04-14T11:28:41},
  doi          = {10.1177/2331216518790903},
  publisher    = {SAGE Publications},
  readstatus   = {skimmed},
}

@Article{Hassager2017,
  author       = {Hassager, Henrik Gert and May, Tobias and Wiinberg, Alan and Dau, Torsten},
  journal      = {The Journal of the Acoustical Society of America},
  title        = {Preserving spatial perception in rooms using direct-sound driven dynamic range compression},
  year         = {2017},
  issn         = {1520-8524},
  month        = jun,
  number       = {6},
  pages        = {4556--4566},
  volume       = {141},
  creationdate = {2025-01-30T14:13:12},
  doi          = {10.1121/1.4984040},
  publisher    = {Acoustical Society of America (ASA)},
  readstatus   = {skimmed},
}

@Article{Zhang2025,
  author       = {Zhang, Huiyong and Moore, Brian C. J. and Jiang, Feng and Diao, Mingfang and Ji, Fei and Li, Xiaodong and Zheng, Chengshi},
  journal      = {Trends in Hearing},
  title        = {Neural-{WDRC}: A Deep Learning Wide Dynamic Range Compression Method Combined With Controllable Noise Reduction for Hearing Aids},
  year         = {2025},
  issn         = {2331-2165},
  month        = jan,
  volume       = {29},
  creationdate = {2025-05-06T14:08:30},
  doi          = {10.1177/23312165241309301},
  publisher    = {SAGE Publications},
}

@article{Schoenenberg2014,
  title={Why are you so slow?--Misattribution of transmission delay to attributes of the conversation partner at the far-end},
  author={Schoenenberg, Katrin and Raake, Alexander and Koeppe, Judith},
  journal={International journal of human-computer studies},
  volume={72},
  number={5},
  pages={477--487},
  year={2014},
  publisher={Elsevier},
  doi={10.1016/j.ijhcs.2014.02.004}
}

@article{Doering2022b,
  author  = {D{\"o}ring, Nicola and Conde, Melisa and Brandenburg, Karlheinz and Broll, Wolfgang and Gross, Horst-Michael and Werner, Stephan and Raake, Alexander},
  title   = {Can communication technologies reduce loneliness and social isolation in older people? A scoping review of reviews},
  journal = {International Journal of Environmental Research and Public Health},
  year    = {2022},
  volume  = {19},
  number  = {18},
  pages   = {11310},
  publisher = {MDPI},
  doi     = {10.3390/ijerph191811310}
}

@article{Raake2022,
  title={Technological factors influencing videoconferencing and zoom fatigue},
  author={Raake, Alexander and Fiedler, Markus and Schoenenberg, Katrin and De Moor, Katrien and D{\"o}ring, Nicola},
  journal={arXiv preprint arXiv:2202.01740},
  year={2022}
}

@article{skowronek2022quality,
  title={Quality of experience in telemeetings and videoconferencing: a comprehensive survey},
  author={Skowronek, Janto and Raake, Alexander and Berndtsson, Gunilla H and Rummukainen, Olli S and Usai, Paolino and Gunkel, Simon NB and Johanson, Mathias and Habets, Emanu{\"e}l AP and Malfait, Ludovic and Lindero, David and others},
  journal={IEEE Access},
  volume={10},
  pages={63885--63931},
  year={2022},
  publisher={IEEE}
}

@incollection{eddy2019technology,
  title={Is technology killing human emotion? How computer-mediated communication compares to face-to-face interactions},
  author={Eddy, Anneli},
  booktitle={Proceedings of {Mensch} und {Computer} 2019},
  pages={527--530},
  year={2019}
}

@book{Raake2007,
  title={Speech quality of VoIP: assessment and prediction},
  author={Raake, Alexander},
  year={2007},
  publisher={John Wiley \& Sons}
}

@inproceedings{Gallardo2018,
  title={Effects of transmitted speech bandwidth on subjective assessments of speaker characteristics},
  author={Gallardo, Laura Fern{\'a}ndez},
  booktitle={2018 Tenth International Conference on Quality of Multimedia Experience (QoMEX)},
  pages={1--5},
  year={2018},
  organization={IEEE}
}

@article{kroczek2021time,
  title={The time course of speaker-specific language processing},
  author={Kroczek, Leon O. H. and Gunter, Thomas C},
  journal={Cortex},
  volume={141},
  pages={311--321},
  year={2021},
  publisher={Elsevier},
  doi={10.1016/j.cortex.2021.04.017}
}

@article{romero2021visual,
  title={Visual information and communication context as modulators of interpersonal coordination in face-to-face and videoconference-based interactions},
  author={Romero, Veronica and Paxton, Alexandra},
  journal={Acta Psychologica},
  volume={221},
  pages={103453},
  year={2021},
  publisher={Elsevier},
  doi ={10.1016/j.actpsy.2021.103453}
}

@ARTICLE{Martin01,
  author={Martin, R.},
  journal={IEEE Transactions on Speech and Audio Processing}, 
  title={Noise power spectral density estimation based on optimal smoothing and minimum statistics}, 
  year={2001},
  volume={9},
  number={5},
  pages={504-512},
  doi={10.1109/89.928915}}

@ARTICLE{Holler15,
  
AUTHOR={Holler, Judith  and Kendrick, Kobin H.  and Casillas, Marisa  and Levinson, Stephen C. },
         
TITLE={Editorial: Turn-Taking in Human Communicative Interaction},
        
JOURNAL={Frontiers in Psychology},
        
VOLUME={Volume 6 - 2015},

YEAR={2015},

URL={https://www.frontiersin.org/journals/psychology/articles/10.3389/fpsyg.2015.01919},

DOI={10.3389/fpsyg.2015.01919},

ISSN={1664-1078},

}

@article{llorca2021multi,
  title={Multi-detailed 3D architectural framework for sound perception research in virtual reality},
  author={Llorca-Bof{\'\i}, Josep and Vorl{\"a}nder, Michael},
  journal={Frontiers in Built Environment},
  volume={7},
  pages={687237},
  year={2021},
  publisher={Frontiers Media SA},
  doi={10.3389/fbuil.2021.687237}
}

@inproceedings{brinkmann_assessing_2014,
  title = {Assessing the Authenticity of Individual Dynamic Binaural Synthesis},
  booktitle = {Proc. of the {{EAA Joint Symp}}. on {{Auralization}} and {{Ambisonics}}},
  author = {Brinkmann, Fabian and Lindau, Alexander and Weinzierl, Stefan},
  date = {2014},
  pages = {62--68},
  location = {Berlin},
  langid = {english}
}

@article{starz_comparison_2025,
  title = {Comparison of Binaural Auralisations to a Real Loudspeaker in an Audiovisual Virtual Classroom Scenario: Effect of Room Acoustic Simulation, {{HRTF}} Dataset, and Head-Mounted Display on Room Acoustic Perception},
  shorttitle = {Comparison of Binaural Auralisations to a Real Loudspeaker in an Audiovisual Virtual Classroom Scenario},
  author = {Stärz, Felix and Par, Steven Van De and Roßkopf, Sarah and Kroczek, Leon O. H. and Mühlberger, Andreas and Blau, Matthias},
  date = {2025},
  journaltitle = {Acta Acustica},
  shortjournal = {Acta Acust.},
  volume = {9},
  pages = {31},
  publisher = {EDP Sciences},
  issn = {2681-4617},
  doi = {10.1051/aacus/2025012},
  url = {https://acta-acustica.edpsciences.org/articles/aacus/abs/2025/01/aacus240088/aacus240088.html},
  urldate = {2025-05-08},
  langid = {english}
}

@book{blauert_technology_2020,
  title = {The {{Technology}} of {{Binaural Understanding}}},
  editor = {Blauert, Jens and Braasch, Jonas},
  date = {2020},
  series = {Modern {{Acoustics}} and {{Signal Processing}}},
  publisher = {Springer International Publishing},
  location = {Cham},
  doi = {10.1007/978-3-030-00386-9},
  url = {http://link.springer.com/10.1007/978-3-030-00386-9},
  urldate = {2025-02-18},
  isbn = {978-3-030-00385-2 978-3-030-00386-9},
  langid = {english}
}

@inproceedings{jaeger_echtzeitfaehiges_2017,
  title = {Echtzeitfähiges Binaurales Rendering mit Bewegungssensoren von 3D-Brillen},
  booktitle = {Fortschritte der Akustik (DAGA)},
  author = {Jaeger, H and Bitzer, J and Simmer, U and Blau, M},
  date = {2017},
  location = {Kiel},
  eventtitle = {DAGA},
  langid = {ngerman}
}

@online{jaeger_timevariant_2023,
  title = {Time-{{Variant Overlap-Add}} in {{Partitions}}},
  author = {Jaeger, Hagen and Simmer, Uwe and Bitzer, Jörg and Blau, Matthias},
  date = {2023},
  url = {http://arxiv.org/abs/2310.00319},
  langid = {english},
  pubstate = {prepublished}
}

@article{grimm2019toolbox,
  title={A toolbox for rendering virtual acoustic environments in the context of audiology},
  author={Grimm, Giso and Luberadzka, Joanna and Hohmann, Volker},
  journal={Acta acustica united with acustica},
  volume={105},
  number={3},
  pages={566--578},
  year={2019},
  publisher={European Acoustics Association}
}

@incollection{pichora2017older,
  title={Older adults at the cocktail party},
  author={Pichora-Fuller, M Kathleen and Alain, Claude and Schneider, Bruce A},
  booktitle={The Auditory System at the Cocktail Party},
  pages={227--259},
  year={2017},
  publisher={Springer}
}

@article{Ermert2025b,
title={Serial recall in spatial acoustic environments: irrelevant sound effect and spatial source alternations}, 
author={Ermert, Cosima A and Yadav, Manuj and Marsh, John E and Schlittmeier, Sabine J and Kuhlen, Torsten W and Fels, Janina}, 
journal={Scientific Reports}, volume={15}, number={1}, pages={32473}, year={2025}, publisher={Nature Publishing Group UK London},
doi = {10.1038/s41598-025-18592-9 } 
}

@article{Seitz2024,
    title={Listening effort in children and adults in classroom noise},
    author={Seitz, J. and Loh, K. and Fels, J.},
    year={2024},
    journal={Sci Rep},
    volume   = {14},
    number   = {2},
    pages={25200},
    doi = {10.1038/s41598-024-76932-7}
}

@article{Yadav2023,
title = {Cognitive performance in open-plan office acoustic simulations: Effects of room acoustics and semantics but not spatial separation of sound sources},
journal = {Applied Acoustics},
volume = {211},
pages = {109559},
year = {2023},
issn = {0003-682X},
doi = {10.1016/j.apacoust.2023.109559},
url = {https://www.sciencedirect.com/science/article/pii/S0003682X23003572},
author = {Manuj Yadav and Markus Georgi and Larissa Leist and Maria Klatte and Sabine J. Schlittmeier and Janina Fels}}

@article{Ermert2025a, title={Audiovisual angle and voice incongruence do not affect audiovisual verbal short-term memory in virtual reality}, author={Ermert, Cosima A and Yadav, Manuj and Ehret, Jonathan and Mohanathasan, Chinthusa and B{\"o}nsch, Andrea and Kuhlen, Torsten W and Schlittmeier, Sabine J and Fels, Janina}, journal={PLoS One}, volume={20}, number={8}, pages={e0330693}, year={2025}, publisher={Public Library of Science San Francisco, CA USA}, doi = {10.1371/journal.pone.0330693 } }

@Article{Breuer2022,
AUTHOR = {Breuer, Carolin and Loh, Karin and Leist, Larissa and Fremerey, Stephan and Raake, Alexander and Klatte, Maria and Fels, Janina},
TITLE = {Examining the Auditory Selective Attention Switch in a Child-Suited Virtual Reality Classroom Environment},
JOURNAL = {International Journal of Environmental Research and Public Health},
VOLUME = {19},
YEAR = {2022},
NUMBER = {24},
ARTICLE-NUMBER = {16569},
URL = {https://www.mdpi.com/1660-4601/19/24/16569},
PubMedID = {36554463},
ISSN = {1660-4601},
DOI = {10.3390/ijerph192416569}
}

@article{Breuer2025,
title={The influence of complex classroom noise on auditory selective attention},
volume={},
DOI={10.1038/s41598-025-18232-2},
journal={Scientific Reports},
author={Carolin Breuer and Robert Josef Schmitt and Larissa Leist and Stephan Fremerey and Alexander Raake and Maria Klatte and Janina Fels},
year={2025},
pages={} 
}

@article{Ermert2025c, author = {Cosima A. Ermert and Karin Loh and Karl Baylan and Konstantin W. K{\"u}hlem and Andrea B{\"o}nsch and Torsten W. Kuhlen and Janina Fels}, title = {Heard-text recall and listening effort under irrelevant speech and pseudo-speech in virtual reality}, journal = {Acta Acustica}, year = {2025}, note = {Manuscript submitted for publication} }

@Article{Oberem2016,
  Title                    = {Experiments on authenticity and plausibility of binaural reproduction via headphones employing different recording methods},
  Author                   = {Josefa Oberem and Bruno Masiero and Janina Fels},
  Journal                  = {Applied Acoustics},
  Year                     = {2016},
  Pages                    = {71 - 78},
  Volume                   = {114},
  Doi                      = {10.1016/j.apacoust.2016.07.009},
  ISSN                     = {0003-682X},
  Url                      = {http://www.sciencedirect.com/science/article/pii/S0003682X16302031}
}

@InProceedings{Masiero2011,
  Title                    = {{Perceptually Robust Headphone Equalization for Binaural Reproduction}},
  Author                   = {Masiero, Bruno S. and Fels, Janina},
  Booktitle                = {Audio Engineering Society Convention 130},
  Year                     = {2011},

  Address                  = {London, United Kingdom},
  Month                    = {13 - 16 May},
  Publisher                = {Red Hook, NY},
  Series                   = {ISBN: 978-0-937809-78-3},
  Url                      = {http://www.aes.org/e-lib/browse.cfm?elib=15855}
}

@Article{Fels2009,
  author    = {Janina Fels and Michael Vorländer},
  title     = {{Anthropometric Parameters Influencing Head-Related Transfer Functions}},
  journal   = {ACTA ACUSTICA united with ACUSTICA},
  year      = {2009},
  volume    = {95},
  number    = {2},
  pages     = {331-342},
  month     = {March/April},
}

@software{institute_for_hearing_technology_and_aco_2024_13788752,
  author       = {
Schäfer, Philipp and Palenda, Pascal and
Aspöck, Lukas and
Fels, Janina and
Vorländer, Michael},
  title        = {Virtual Acoustics - A real-time auralization
 framework for scientific research
},
  year         = 2024,
  publisher    = {Zenodo},
  version      = {2024a},
  doi          = {10.5281/zenodo.13788752},
  url          = {https://doi.org/10.5281/zenodo.13788752},
address = {{Institute for Hearing Technology and Acoustics, RWTH Aachen University}}
}

@inproceedings{Ehret2020i,
author = {Ehret, Jonathan and Stienen, Jonas and Brozdowski, Chris and B\"{o}nsch, Andrea and Mittelberg, Irene and Vorl\"{a}nder, Michael and Kuhlen, Torsten W.},
title = {Evaluating the Influence of Phoneme-Dependent Dynamic Speaker Directivity of Embodied Conversational Agents' Speech},
year = {2020},
isbn = {9781450375863},
publisher = {Association for Computing Machinery},
address = {New York, NY, USA},
url = {https://doi.org/10.1145/3383652.3423863},
doi = {10.1145/3383652.3423863},
booktitle = {Proceedings of the 20th ACM International Conference on Intelligent Virtual Agents},
articleno = {17},
numpages = {8},
location = {Virtual Event, Scotland, UK},
series = {IVA '20}
}

@Article{Rubio-Tamayo2017immersive,
AUTHOR = {Rubio-Tamayo, Jose Luis and Gertrudix Barrio, Manuel and García García, Francisco},
TITLE = {Immersive Environments and Virtual Reality: Systematic Review and Advances in Communication, Interaction and Simulation},
JOURNAL = {Multimodal Technologies and Interaction},
VOLUME = {1},
YEAR = {2017},
NUMBER = {4},
ARTICLE-NUMBER = {21},
URL = {https://www.mdpi.com/2414-4088/1/4/21},
ISSN = {2414-4088},
DOI = {10.3390/mti1040021}
}

@inproceedings{Aspoeck2014,
  author    = {Asp{\"o}ck, Lukas and Pelzer, S{\"o}nke and Wefers, Frank and Vorl{\"a}nder, Michael},
  title     = {A real-time auralization plugin for architectural design and education},
  booktitle = {Proceedings of the EAA Joint Symposium on Auralization and Ambisonics},
  year      = {2014},
  pages     = {156--161},
  address   = {Berlin, Germany},
  month     = apr,
  doi       = {10.14279/depositonce-26},
}

@Article{Nirme.2018,
  author   = {Nirme, Jens and Haake, Magnus and Lyberg {\AA}hlander, Viveka and Br{\"a}nnstr{\"o}m, Jonas and Sahl{\'e}n, Birgitta},
  title    = {A virtual speaker in noisy classroom conditions: Supporting or disrupting children's listening comprehension?},
  journal  = {Logopedics Phoniatrics Vocology},
  year     = {2018},
  pages    = {79--86},
  doi      = {10.1080/14015439.2018.1455894}
}

@Article{Oberem2018a,
  author  = {Oberem, Josefa and Seibold, Julia and Koch, Iring and Fels, Janina},
  title   = {Intentional switching in auditory selective attention: Exploring attention shifts with different reverberation times},
  journal = {Hearing Research},
  year    = {2018},
  volume  = {359},
  pages   = {32--39},
  issn    = {0378-5955},
  doi     = {10.1016/j.heares.2017.12.013}
}

@Article{shinn2008object,
  author    = {Shinn-Cunningham, Barbara G},
  title     = {Object-based auditory and visual attention},
  journal   = {Trends Cogn Sci.},
  year      = {2008},
  volume    = {12},
  number    = {5},
  pages     = {182--186},
  publisher = {Elsevier},
}

@Article{devesse.2018,
  author   = {Devesse, Annelies and Dudek, Alexander and {van Wieringen}, Astrid and Wouters, Jan},
  title    = {Speech intelligibility of virtual humans},
  journal  = {International Journal of Audiology},
  year     = {2018},
  volume={57},
  number={12},
  pages    = {1--9},
}

@article{Brady1965,
  title={A technique for investigating on-off patterns of speech},
  author={Brady, Paul T},
  journal={Bell System Technical Journal},
  volume={44},
  number={1},
  pages={1--22},
  year={1965},
  publisher={Wiley Online Library}
}

@book{mollerAssessmentPredictionSpeech2000,
  title = {Assessment and {{Prediction}} of {{Speech Quality}} in {{Telecommunications}}},
  author = {Möller, Sebastian},
  date = {2000},
  publisher = {Springer US},
  location = {Boston, MA},
  doi = {10.1007/978-1-4757-3117-0},
  url = {http://link.springer.com/10.1007/978-1-4757-3117-0},
  urldate = {2025-11-14},
  isbn = {978-1-4419-4989-9 978-1-4757-3117-0},
  langid = {english}
}

@INPROCEEDINGS{Liebich18,
  author={Liebich, Stefan and Fabry, Johannes and Jax, Peter and Vary, Peter},
  booktitle={Speech Communication; 13th ITG-Symposium}, 
  title={Signal Processing Challenges for Active Noise Cancellation Headphones}, 
  year={2018},
  volume={},
  number={},
  pages={1-5},
  doi={}}

@inproceedings{dehaas_realtime_2025,
  title = {Real-Time Virtual Environment and Room Acoustics Simulator},
  booktitle = {Proc. {{Forum Acusticum}}},
  author = {family=Haas, given=Kevin, prefix=de, useprefix=true and Schutte, Michael and Ewert, Stephan D},
  date = {2025},
  location = {Málaga},
  eventtitle = {Forum {{Acusticum Euronoise}}},
  langid = {english}
}

@book{xie2013head,
  title={Head-related transfer function and virtual auditory display},
  author={Xie, Bosun},
  year={2013},
  publisher={J. Ross Publishing}
}

@book{wefers_partitioned_2015,
  author    = {Wefers, Frank},
  title     = {Partitioned convolution algorithms for real-time auralization},
  date      = {2015},
  publisher = {Logos Verlag Berlin GmbH},
  location  = {Berlin},
  langid    = {english},
  pagetotal = {258}
}


\end{document}